\documentclass[3p,times]{elsarticle}

\usepackage{ecrc}

\volume{00}

\firstpage{1}

\journalname{Annual Reviews in Control}

\runauth{}

\jid{procs}

\jnltitlelogo{Annual Reviews in Control}

\CopyrightLine{2022}{Published by Elsevier Ltd.}

\usepackage{amsmath,amssymb}
\usepackage{subfigure}
\usepackage{comment,xcolor}
\usepackage{subfigmat}
\usepackage[figuresright]{rotating}

\newcommand{\R}{\ensuremath{{\mathbb R}}}
\newcommand{\Z}{\ensuremath{{\mathbb Z}}}
\newcommand{\N}{\ensuremath{{\mathbb N}}}

\newcommand{\ra}{\rightarrow}
\newcommand{\xxx}{\mathsf{x}}
\newcommand{\yyy}{\mathsf{y}}
\newcommand{\uuu}{\mathsf{u}}
\newcommand{\zzz}{\mathsf{z}}

\newcommand{\mmm}{\mathsf{m}}
\newcommand{\ppp}{\mathsf{p}}

\newcommand{\rrr}{\mathsf{r}}
\newcommand{\sss}{\mathsf{s}}
\newcommand{\LLL}{\mathsf{L}}
\newcommand{\RRR}{\mathsf{R}}

\newcommand{\FFF}{\mathsf{F}}
\newcommand{\GGG}{\mathsf{G}}
\newcommand{\HHH}{\mathsf{H}}
\newcommand{\JJJ}{\mathsf{J}}

\newcommand{\FF}{\mathbf{F}}
\newcommand{\GG}{\mathbf{G}}
\newcommand{\HH}{\mathbf{H}}
\newcommand{\JJ}{\mathbf{J}}

\newcommand{\RR}{\mathbf{R}}

\newcommand{\xx}{\mathbf{x}}
\newcommand{\zz}{\mathbf{z}}
\newcommand{\uu}{\mathbf{u}}
\newcommand{\yy}{\mathbf{y}}

\newcommand{\cc}{\mathbf{c}}
\newcommand{\CC}{\mathbf{C}}

\newcommand{\ol}{\overline}
\newcommand{\col}{\mathrm{col}}

\newcommand{\CCCC}{{\mathcal C}}

\newcommand{\Enc}{\mathsf{Enc}}
\newcommand{\Dec}{\mathsf{Dec}}

\newcommand{\Mult}{\mathsf{Mult}}

\newcommand{\sk}{\mathsf{sk}}

\colorlet{Darkred}{red!50!black}
\colorlet{Darkgreen}{green!50!black}
\colorlet{Darkblue}{blue!70!black}
\colorlet{ggray}{white!90!black}

\newcommand{\colg}{\color{Darkgreen}}
\newcommand{\colr}{\color{red}}
\newcommand{\colo}{\color{orange}}

\newtheorem{thm1}{\bf Theorem}
\newtheorem{prop1}{\bf Proposition}
\newtheorem{lem1}{\bf Lemma}
\newtheorem{asm1}{\bf Assumption}
\newtheorem{defn1}{\bf Definition}
\newtheorem{rem1}{\bf Remark}
\newtheorem{cor1}{\bf Corollary}

\newtheorem{proof1}{\it Proof}
\newtheorem{prob1}{\bf Problem}

\newenvironment{asm}{\begin{asm1}}{\hfill$\square$\end{asm1}}

\newenvironment{lem}{\begin{lem1}}{\hfill$\square$\end{lem1}}
\newenvironment{thm}{\begin{thm1}}{\hfill$\square$\end{thm1}}
\newenvironment{cor}{\begin{cor1}}{\hfill$\square$\end{cor1}}
\newenvironment{prop}{\begin{prop1}}{\hfill$\square$\end{prop1}}
\newenvironment{prob}{\begin{prob1}}{\hfill$\square$\end{prob1}}
\begin{document}
	
	\begin{frontmatter}
		
		

		\dochead{}
		%% e.g. \dochead{17th International Conference on Dynamical Processes in Excited States of Solids}
		
		\title{
			%Encrypted control approaches and methods for dynamic systems based on homomorphic encryption
			Comparison of encrypted control approaches and tutorial on dynamic systems using LWE-based homomorphic encryption
		}
		

		\author[First]{\normalsize Junsoo Kim} 
		\author[Second]{Dongwoo Kim} 
		\author[Third]{Yongsoo Song}
		\author[Fourth]{Hyungbo Shim}
		\author[Fifth]{Henrik Sandberg}
		\author[Fifth]{Karl H. Johansson}
		
		\address[First]{Department of Electrical and Information Engineering, Seoul National University of Science and Technology, Seoul, Korea}
		\address[Second]{Western Digital Research, Milpitas, California, USA}
		\address[Third]{
	Department of Computer Science and Engineering, Seoul National University, Seoul, Korea
	}
		\address[Fourth]{ASRI, Department of Electrical and Computer Engineering,
			Seoul National University, Seoul, Korea}
		\address[Fifth]{School of Electrical Engineering and Computer Science and Digital Futures, KTH Royal Institute of Technology, Stockholm, Sweden}

		\begin{abstract}
			Encrypted control has been introduced to protect controller data by encryption at the stage of computation and communication, by performing the computation directly on encrypted data. In this article, we first review and categorize recent relevant studies on encrypted control. Approaches based on homomorphic encryption, multi-party computation, and secret sharing are introduced, compared, and then discussed with respect to computational complexity, communication load, enabled operations, security, and research directions. We proceed to discuss a current challenge in the application of homomorphic encryption  to dynamic systems, where arithmetic operations other than integer addition and multiplication are limited. We also introduce a homomorphic cryptosystem called ``GSW-LWE'' and discuss its benefits that allow for recursive multiplication of encrypted dynamic systems, without use of computationally expensive bootstrapping techniques.
		\end{abstract}
		
		\begin{keyword}
			
			
			Encrypted control, homomorphic encryption, bootstrapping, multi-party computation, secret sharing, dynamic system over encrypted data, Learning With Errors
		\end{keyword}
		
	\end{frontmatter}
	
	%%\linenumbers
	
	\section{Introduction}\label{sec:intro}

	Networked control has enabled significant developments in numerous industrial fields, while it has also led to urgent issues related to cyber-security \cite{sandbergCSM15,aminHSCC09}.
	The more control systems have been connected to networks, the more  possibilities for cyber-attacks have been discovered. Given the inherent cyber-physical nature of these systems, such attacks risk also the safe operation of the connected physical systems.
	Many incidents have been reported in recent years, including the StuxNet worm \cite{langner11}, false data injection to power grids \cite{liu11}, and a security breach in water sewage system \cite{slay07}.

\begin{figure}[t]
		\centering
		\includegraphics[width=.5\columnwidth]{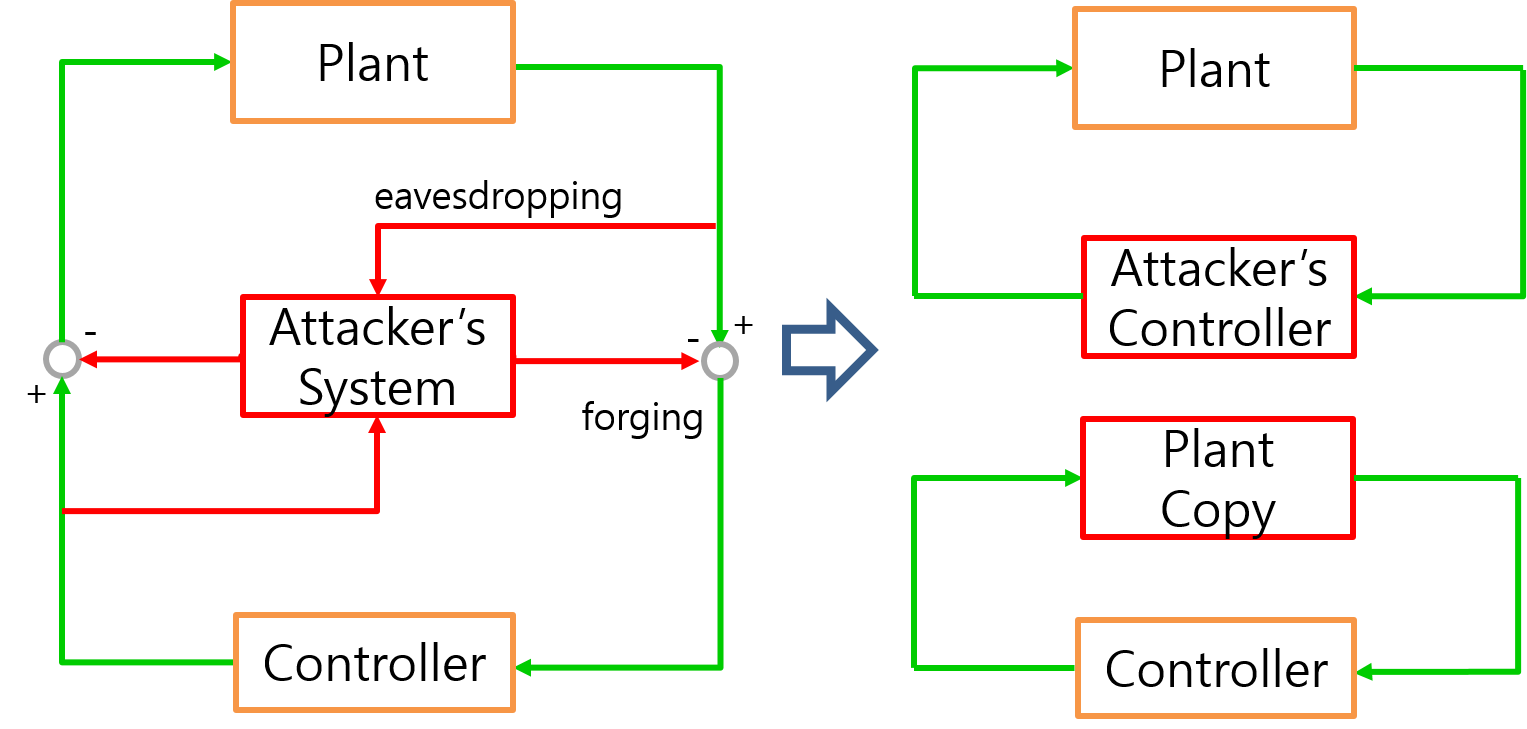}
		\caption{A possible scenario \cite{smith15CSM} when all the information of system model and control signals is used for cyber-attacks.}
		\label{fig:covert}
	\end{figure}

	Studies in security of control systems are distinguished from those in fault detection and isolation in the sense that the considered attacks are malicious, not random faults, and may be elaborately designed to avoid detection \cite{teixeira15AUT}.
	A negligible probability of detection of a complex fault may be acceptable in safety studies, but in security such cases can be the point where the attacker deceives the detector.
	For example, zero-dynamics attacks \cite{pasqualetti13TAC} exploiting  system models, replay attacks \cite{mo09} exploiting signals, and covert attacks \cite{smith15CSM} exploiting both models and signals, have been reported.
	A common point in these attacks is that {\it the more the attacker learns of the target system, the more effective and theoretically undetectable attacks can be designed}.
	An extreme case is the covert attack presented in \cite{smith15CSM};
	if an adversary
	can compromise both the input and output communication using all the information of model and signals,
	then it can decouple the closed-loop of the plant and the controller, and manipulate the plant arbitrarily. This is while the controller (with any anomaly detector) cannot distinguish whether the received signals are compromised, or not (see Figure~\ref{fig:covert}).
	
\begin{figure}[b]
		\centering
		\begin{subfigmatrix}{4}
			\subfigure[]{\includegraphics[width=.4\textwidth]{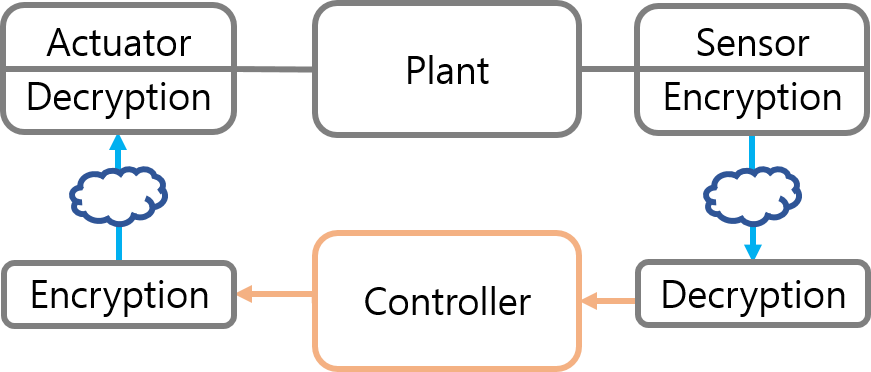}\label{fig:config1}}		
			\subfigure[]{\includegraphics[width=.4\textwidth,keepaspectratio]{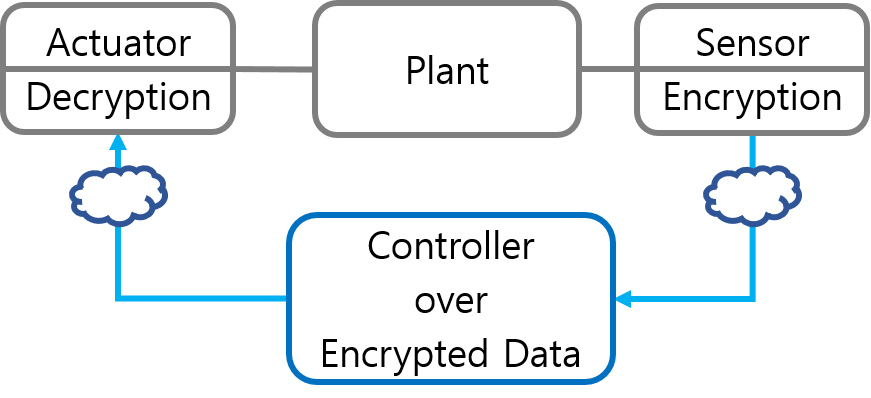}\label{fig:config2}}		
		\end{subfigmatrix}
		\caption{(a) Data protection by conventional encryption. (b) Configuration of encrypted control.
		}
		\label{fig:config}
	\end{figure}

	From the motivation that
	data stored or being transmitted in networks may be used for advanced attack generation,
	the notion of {\it encrypted control} has been introduced \cite{darup21CSM,KogisoCDC15,kim16NECSYS,farokhi17CEP}, which aims for security enhancement by protecting all data in networked controllers by encryption.
	Conventional encryption has been used in data transmission, as illustrated in Figure~\ref{fig:config1}),
	but computing devices in the network layer have been regarded as one of the most vulnerable parts.
	Since cryptosystems allowing arithmetic operations was invented \cite{rivest78}, more developed schemes \cite{elgamal85,paillier99,cheon17} have been applied to control as in \cite{KogisoCDC15,kim16NECSYS,farokhi17CEP}. The concept of ``control operation over encrypted data,'' as illustrated in Figure~\ref{fig:config2}, is therefore a promising direction for improving cyber-security.
	
	Benefits of the solution in Figure~\ref{fig:config2} are that all data can be kept secure even when computation is performed,
	without decryption, and the secret key for decryption is not shared.
	Unauthorized parties, or possible network intruders, cannot learn any information through eavesdropping as long as they do not have the secret key, because encrypted signals cannot be distinguished from uniformly generated random signals.
	Furthermore, a typical type of cryptosystem used for encrypted operation, called homomorphic encryption, is known to be more beneficial than other security mechanisms.
	For example, compared to methods based on
	differential privacy \cite{dwork08,dwork14}, it does not sacrifice precision of the encrypted messages, regardless of the increase of the security level. Also, many homomorphic encryption schemes are based on discrete lattice problems, such as \cite{regev2009,lyubashevsky10}, and consider a stronger adversarial model than other schemes, such as chaotic cryptology techniques \cite{matthews89,bourbakis92}.

	Nonetheless,
	the more operational abilities in recent cryptosystems are generally accompanied by increased costs in terms of computation or communication resource use. From an engineering point of view,
	application of cryptosystems to control thus should consider the following aspects for design.
	
	\begin{itemize}
		\item {\it Type of controller, required operations, and available resources}.
		Cryptosystems that allow arithmetic operations or guarantee a higher level of security usually require more data storage for encrypted data.
		The computational effort or communication overhead taken for each use of the arithmetic function may cost much more than usual operation over un-encrypted messages. Possibly, only one type of arithmetic function is supported, or only a limited number of function evaluations may be allowed.
		Thus, the design of encrypted control systems should consider
		the required type of operations and their complexity; the
		available amount of computation and communication resources;
		the time use for each arithmetic computation;
		whether the system is static or dynamic;
		the available amount of data storage; and if
		there are multiple units cooperating in the computation or
		there is only a single unit.
		
		\item {\it Security model (adversarial model)}.
		The design should include an adversarial model, which determines from what attacks the system is protected. The adversary of interest can be an unauthorized party in the network, or an external hacker intruding the network.
		The adversary may aim for learning certain information by eavesdropping the communication lines or networked devices;
		or for manipulating or forging a portion of data with a certain objective in mind.
		Depending on the number of units and the structure of the design, the security objective can be to protect the data from external parties only, or to conceal it from participating units who may be curious about other participants' private information.

	\end{itemize}
	
	Considering the above aspects, the designer of an encrypted control system chooses an appropriate cryptosystem, which may be computationally fast with only limited arithmetic support, may have much more operation abilities with requirement of large amounts of resources, or can be suitable for centralized or distributed computation schemes, respectively. Usually, there are trade-offs between the cryptosystems, in terms of use of resources, level of security, or enabled encrypted operations.
	Relevant research efforts try to overcome this trade-off; by re-constructing a computation system which exploits only a limited type of operation,	by developing a novel cryptosystem that is more appropriate for a certain application, or by proposing a protocol that guarantees an improved level of security. In short, the efforts try to improve the security or computational efficiency, while not losing operational abilities.
	
	\medskip
	
	{\it Contribution and organization:}
	This article provides both a general review of the field of encrypted control and a comprehensive introduction of the specific topic of dynamic system design based on homomorphic encryption.
	In Section~\ref{sec:encrypted_control},
	we discuss main approaches for encrypted control, which are based on {\it homomorphic encryption, multi-party computation, or secret sharing}, respectively.
	These cryptosystems, as tools for encrypted control, are briefly introduced, and the approaches and their application to control are compared. Trade-offs between computation speed, operation ability, and security are also discussed.
	In Section~\ref{sec:dynamic},
	we focus on the homomorphic encryption based approach
	and introduce recent methods for linear dynamic systems.
	A tutorial on a Learning With Errors (LWE) based homomorphic encryption scheme is first provided, and we discuss its benefit in terms of recursive multiplication.
	Then, we review a method that implements linear systems by exploiting only the addition and multiplication abilities of cryptosystem.
	Finally, Section~\ref{sec:conclu} concludes the paper.

	{\it Notation:}
	Let $\N$, $\Z$, and $\R$ denote the set of natural numbers, integers,  and real numbers, respectively.
	The (component-wise) floor and round functions are denoted by $\lfloor\cdot\rfloor$ and $\lceil \cdot \rfloor$, respectively.
	For $m\in\N$ and $n\in\N$, we let $0_{m\times n}\in\R^{m\times n}$ be the zero matrix,
	and $I_n\in\R^{n\times n}$ be the identity matrix.
	The set of integers modulo $q\in\N$ is denoted by
	$\Z_q$, and the (component-wise) modulo operation is defined by $v \mod q:= v- \lfloor v/q\rfloor q$ for $v\in\Z^m$.
	We make use of ``biased'' modulo operation defined as \begin{equation}\label{eq:biased_modulo}
		v\mod (q,v_0) := v- \left\lfloor \frac{v-v_0}{q}\right\rfloor q
	\end{equation}
	for $v\in\Z^m$ and $v_0\in\R^m$,
	so that each component of the outcome is greater than or equal to that of $v_0$, and less than that of $v_0 + q$.
	We define $\col\{v_i\}_{i=1}^{n}=[v_1^\top,\cdots,v_n^\top]^\top$ for column vectors $\{v_i\}_{i=1}^{n}$ (or scalars),
	and the (induced) infinity norm of a vector or a matrix is denoted as $\|\cdot\|$.
	
	\section{Encrypted control approaches}\label{sec:encrypted_control}
	
	This section categorizes the relevant works into four categories according to the underlying cryptographic primitives; homomorphic encryption, fully homomorphic encryption, secret sharing, and multi-party computation, together with a brief introduction to each primitive and approach.
	Then, we provide discussions and comparisons between them.

	\subsection{Homomorphic encryption based control}\label{subsec:HE}
	
	We begin with a brief introduction on homomorphic encryption.
	Let a cryptosystem be denoted by $(\Z_q,\CCCC,\Enc,\Dec)$,
	where the set $\Z_q=\{0,1,\ldots,q-1\}$, $q\in\N$, is the plaintext (un-encrypted message) space, $\CCCC$ is the ciphertext (encrypted) space, $\Enc:\Z_q \ra \CCCC$ and $\Dec:\CCCC\ra\Z_q$ are the encryption and decryption algorithms, respectively. We omit the argument of encryption and decryption keys, for simplicity.
	
	Homomorphic properties of cryptosystems imply that the encryption and decryption algorithms are ``homomorphisms,'' which preserve algebraic structure with respect to a certain arithmetic function.
	Let us assume that the cryptosystem $(\Z_q,\CCCC,\Enc,\Dec)$ is additively homomorphic, which means that the algorithms $\Enc$ and $\Dec$ are homomorphic with respect to the addition operation;
	there exists a binary function $\ast:\CCCC\times\CCCC\ra \CCCC$ over ciphertexts such that
	\begin{equation*}
		\Dec(\cc_1\ast \cc_2) = \Dec(\cc_1)+\Dec(\cc_2)\mod q,\qquad \forall \cc_1\in\CCCC,~\forall \cc_2\in\CCCC,
	\end{equation*}
	which implies that
	\begin{equation}\label{eq:additive}
		\cc = \Enc(x_1)\ast \Enc(x_2) \quad \implies \quad \Dec(\cc) = x_1 + x_2\mod q,\quad \forall x_1\in\Z_q,~\forall x_2\in\Z_q.
	\end{equation}
	It means that whenever we decrypt the outcome of the operation $\ast$ over ciphertexts, we obtain the same addition outcome over plaintexts, so it enables to perform the addition directly over encrypted messages, without decryption.
	
	The ability of addition is a basic property, but by exploiting it, multiplication by (un-encrypted) constant numbers
	can also be performed.
	Given a natural number $k\in\N$, we define
	\begin{equation*}
		k\cdot \cc :=  \smash[b]{\overbrace{\cc\ast\cdots\ast \cc}^{\text{$k$ times}}}.
	\end{equation*}
	And, we extend the definition for an integer
	$k\in\Z$,
	by
	$$ k \cdot \cc := (k\!\!\!\mod (q,1))\cdot \cc,$$
	where
	the operation $\mathrm{mod}(q,1)$ defined from \eqref{eq:biased_modulo} ensures that $ k~\mathrm{mod}(q,1)\ge 1$.
	Then, it follows that
	\begin{align*}
		\begin{split}
			\Dec(k\cdot \cc) &= (k\!\!\!\mod (q,1))\cdot \Dec(\cc)\mod q\\
			&=k\cdot \Dec(\cc)\mod q
			\end{split}	
	\end{align*}
	holds
	for all $k\in\Z$ and $\cc\in\CCCC$,
	because
	$k~\mathrm{mod}(q,1)=k + dq$ with some $d\in\Z$.

	It enables multiplication by integer matrices as well; for an integer matrix $K=[k_{ij}]\in\Z^{m\times n}$ and a set $\xx = \{\Enc(x_i)\}_{i=1}^{n}$ of encrypted messages of $x_i\in\Z_q$, $i=1,\ldots,n$, we define
	\begin{equation}\label{eq:matrix_mult}
		K\cdot \xx = \begin{bmatrix}
			k_{11}&k_{12} & \cdots & k_{1 n}\\
			k_{21} & k_{22} & \cdots & k_{2n}\\
			\vdots & \vdots& \ddots & \vdots\\
			k_{m 1} &k_{m2}& \cdots & k_{m n}
		\end{bmatrix}\cdot
		\begin{bmatrix}
			\Enc(x_1)\\ \Enc(x_2)\\ \vdots \\ \Enc(x_n)
		\end{bmatrix}
		:=
		\begin{bmatrix}
			(k_{11}\cdot \Enc(x_1))\ast (k_{12}\cdot \Enc(x_2))\ast\cdots\ast (k_{1n}\cdot \Enc(x_n))\\
			(k_{21}\cdot \Enc(x_1))\ast(k_{22}\cdot \Enc(x_2))\ast\cdots\ast (k_{2n}\cdot \Enc(x_n))\\
			\vdots\\
			(k_{m1}\cdot \Enc(x_1))\ast(k_{m2}\cdot \Enc(x_2))\ast\cdots\ast (k_{mn}\cdot \Enc(x_n))
		\end{bmatrix},
	\end{equation}
	which is simply a component-wise constant multiplication and consecutive additions over encrypted data.
	Then, it can also be easily verified that
	$\Dec(K\cdot \xx) = K\cdot [x_1,x_2,\ldots,x_n]^\top\mod q$, for all $K\in\Z^{m\times n}$ and $x\in\Z_q^n$.
	
	Thanks to the property of allowing computation over ciphertexts without the secret key nor decryption, the computation can be assigned to an operation unit who is honest to do the computation correctly while being curious about the transmitted data, or to a networked unit which can be possibly accessed by unauthorized third parties.
	In addition to 	the described homomorphic properties, there are also encryption schemes~\cite{cheon17,boneh2005evaluating,dijk2010fully,brakerski2011fully,fan12,brakerski14} which enable to compute both the addition and a limited number of multiplication on encrypted data. Those schemes are sometimes called somewhat or leveled homomorphic encryption to be distinguished from (partially) homomorphic encryptions allowing addition only, but we denote both of them, simply, as homomorphic encryption.
	
\begin{figure}[h]
			\centering
			\begin{subfigmatrix}{4}
				\subfigure[]{\includegraphics[width=.46\textwidth]{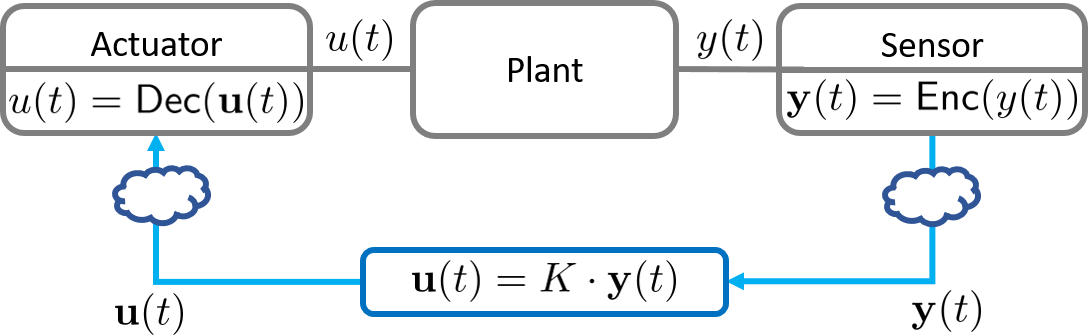}\label{fig:static}}
				\subfigure[]{\includegraphics[width=.46\textwidth,keepaspectratio]{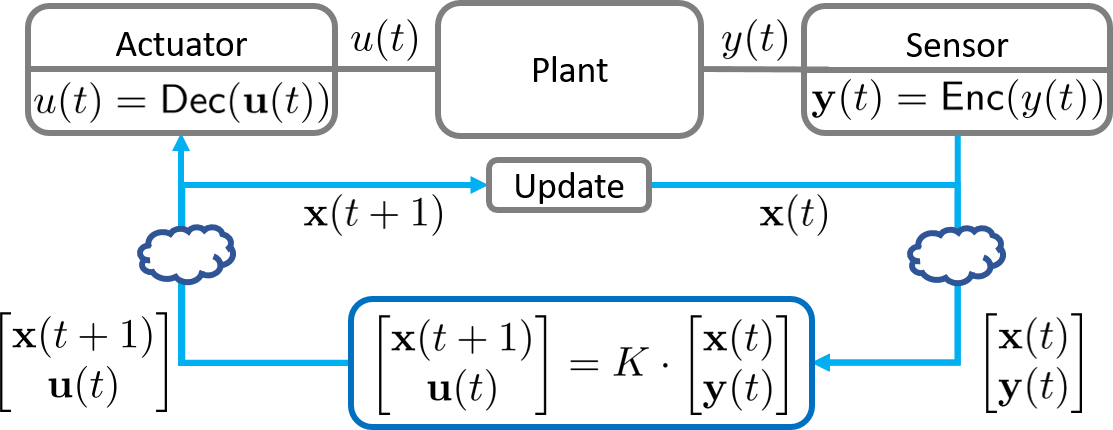}\label{fig:dynamic}}
			\end{subfigmatrix}
			\caption{
				(a) Homomorphic encryption based controllers for static feedback. (b) Dynamic encrypted controller implementation using additional transmission lines for the state.
			}
			\label{fig:HE}
		\end{figure}

	The application of homomorphic encryption for control has been introduced in  \cite{KogisoCDC15,kim16NECSYS,farokhi17CEP}.
	Since homomorphic cryptosystems support addition and/or multiplication, most controllers based on homomorphic encryption considers linear operation (matrix multiplication) or polynomial functions (represented with small number of additions and multiplications) only; see Figure~\ref{fig:static} for an example case of linear controllers based on homomorphic encryption.
	It has also been used for implementing
	model predictive control \cite{darup17CSL},
	data-driven control~\cite{ATP20}, and reinforcement learning based control \cite{suh21acc},
	where the computation circuits are represented with linear functions or low degree polynomials.

		{\it Operation unit and attack model:}
		The model of the controller is usually considered as a single operation unit who performs the computation over encrypted data without accessing the secret key.
		It is often specified as a networked controller under possible eavesdropping attack (which tries to learn private control data) or considered as a cloud-based controller performing the computation on behalf of the controller designer, which should not learn any private information through the computation.
		In short, adversaries of homomorphic encryption based controllers can be both external eavesdroppers and the controller itself, and cryptosystem parameters are determined for a desirable level of security.

	{\it Challenges:}
	A main issue is that encrypted implementation may be limited to only addition or multiplication over integers.
	Even for homomorphic encryption allowing both addition and multiplication, many of them allow evaluating a limited number of multiplications only.
	As a consequence, implementation of dynamic systems that iteratively compute and update the state has been considered as a challenge, and representing diverse nonlinear systems using only addition and multiplication is currently an open problem.

	Even for linear systems,
	dynamic systems implemented over encrypted data may be incapable of operating for an infinite time horizon.
	This is because,
	based on homomorphic encryption only, multiplication by non-integer numbers can be performed only a finite number of times,
	so that
	the dynamic operation that multiplies the state by non-integer numbers cannot be continued for an infinite time horizon (see the example described in \eqref{eq:example}).
	Figure~\ref{fig:dynamic} describes a method for dynamic controllers considered in initial studies, which uses additional transmission lines for the state;
	assuming that the encrypted state of the controller can be transmitted to the device having the decryption key, the state can be decrypted and re-encrypted and transmitted back to the controller. 
	However, the limitation is that the system cannot continue the operation without the presence of the decryption key, and the re-encryption requires additional use of communication lines which may be proportional to the dimension of the signal. 
	Further discussions and related methods can be found in Section~\ref{subsec:method}.

	\subsection{Fully homomorphic encryption based methods}
	
	As introduced, homomorphic encryption supporting a finite number of operations has limitations on its applicability.
	The first fully homomorphic encryption,
	with which one can perform arbitrary computation without limitation,
	has been presented in \cite{gentry09}.
	Then, the main concern has been reducing the computational cost, and follow-up works have been proposed toward their practical use~\cite{cheon17,fan12,brakerski14,ducas15,chillotti16}, with several library implementations.
	The crucial improvement from the previous homomorphic encryption (with limitations) is the introduction of ``bootstrapping'' procedure\footnote{Conceptually, it can be understood as an encrypted evaluation of the decryption operation on the ciphertext, resulting in a new ciphertext having the same message.}~\cite{ducas15,gentry12} which refreshes (without a secret key) a ciphertext into a new one with which one can continue the computation.
	Many encryption schemes referred to in the previous section can be fully homomorphic, by adding the bootstrapping procedure.
	
	\begin{figure}[t]
		\centering
		\includegraphics[width=.46\textwidth]{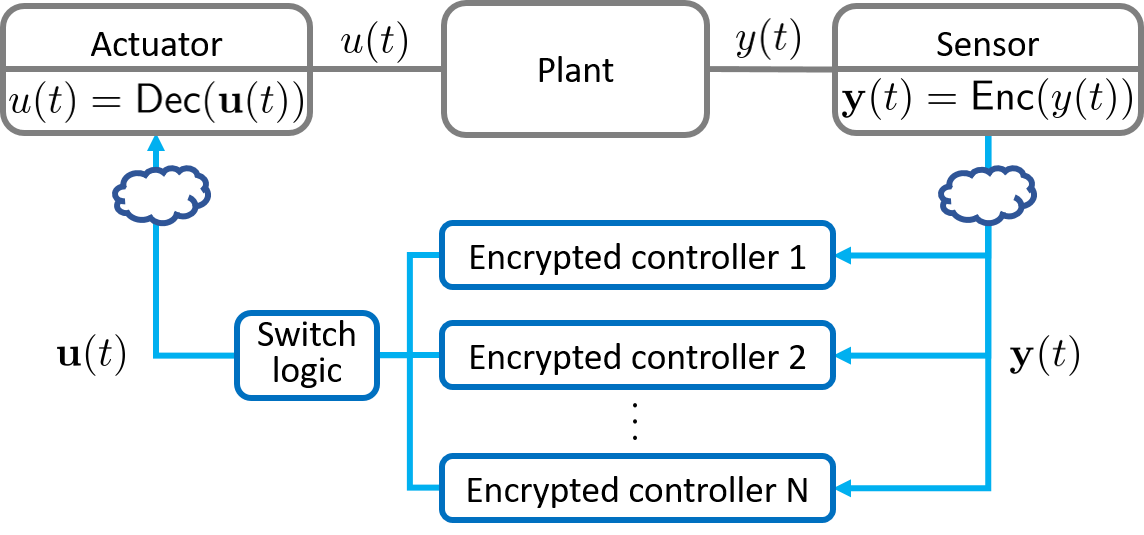}
		\caption{
			Multiple encrypted linear controllers with switching function.
		}
		\label{fig:switch}
	\end{figure}
	
	Thanks to the bootstrapping,
	any operation circuits in digital computers can be implemented and run over encrypted data for an infinite time horizon,
	since any logical functions with Boolean variables can be represented using addition and multiplication.
	Nonetheless,
	the {\it computational complexity of bootstrapping has been a critical issue},
	and its requirement on the computation resource currently hinders it from being used in practice.

	The bootstrapping was first used for linear dynamic systems in \cite{kim16NECSYS}.
	The latency time for bootstrapping may be larger than the sampling period,
	but as described in Figure~\ref{fig:switch},
	multiple identical controllers are designed and the time for bootstrapping is scheduled in a different way, so that the same output can be generated all the time even when some of them are performing bootstrapping.
	Many homomorphic encryption based works~\cite{ATP20,suh21acc,FFZ2019,zhou2020homomorphic} mentioned in the previous section also claim that introduction of bootstrapping enables the encrypted systems to update the state without interacting with an entity holding the secret key.

	{\it Operation unit and attack model:}
	Since fully homomorphic encryption is simply a homomorphic encryption with further evaluation capabilities, the operation model considered is the same as the case of homomorphic encryption without use of bootstrapping.
	It also considers a single operation unit who should not learn any information from the computation, while all the control data are kept encrypted so that it is protected from external parties as well.
	A difference in practice is that it should be able to use a large amount of computational or memory resources, for bootstrapping.
	
	{\it Challenges:}
	Main challenge of fully homomorphic encryptions is to reduce the computational cost for bootstrapping.
	A couple of issues regarding control operation can be listed;
	first, operations consisting of addition, multiplication, and bootstrapping,
	may not be efficient for non-polynomial functions, such as,	
	comparison, if-else conditionals, or transcendental functions,
	for which lots of arithmetic operations may be required for the representation and result in impractical evaluation cost.
	A countermeasure would be to represent the given operation circuit to an ``arithmetic-friendly form,'' so that the evaluation cost can be reduced.
	And, since bootstrapping constitutes a major cost in fully homomorphic encryption based schemes while it has been essentially used, a momentous challenge will be to reduce the number of uses of bootstrapping.

	\subsection{Secret sharing based schemes}\label{subsec:SS}
	
	Now, we describe another computation protocol based on secret sharing, where private data are distributed in the form of ``shares'' to two or more parties so that an individual party cannot get any information about the data.
	Still, each party can perform appropriate computation on its share and generate the outcome from which the owner of the data can reconstruct the computation outcome;
	in other words, there is a homomorphic property on the shares,	so that the computation of control signals can be offloaded to outsourced operation units, without disclosing any information on the input and output.
	
\begin{figure}[h]
		\centering
		\includegraphics[width=.7\textwidth]{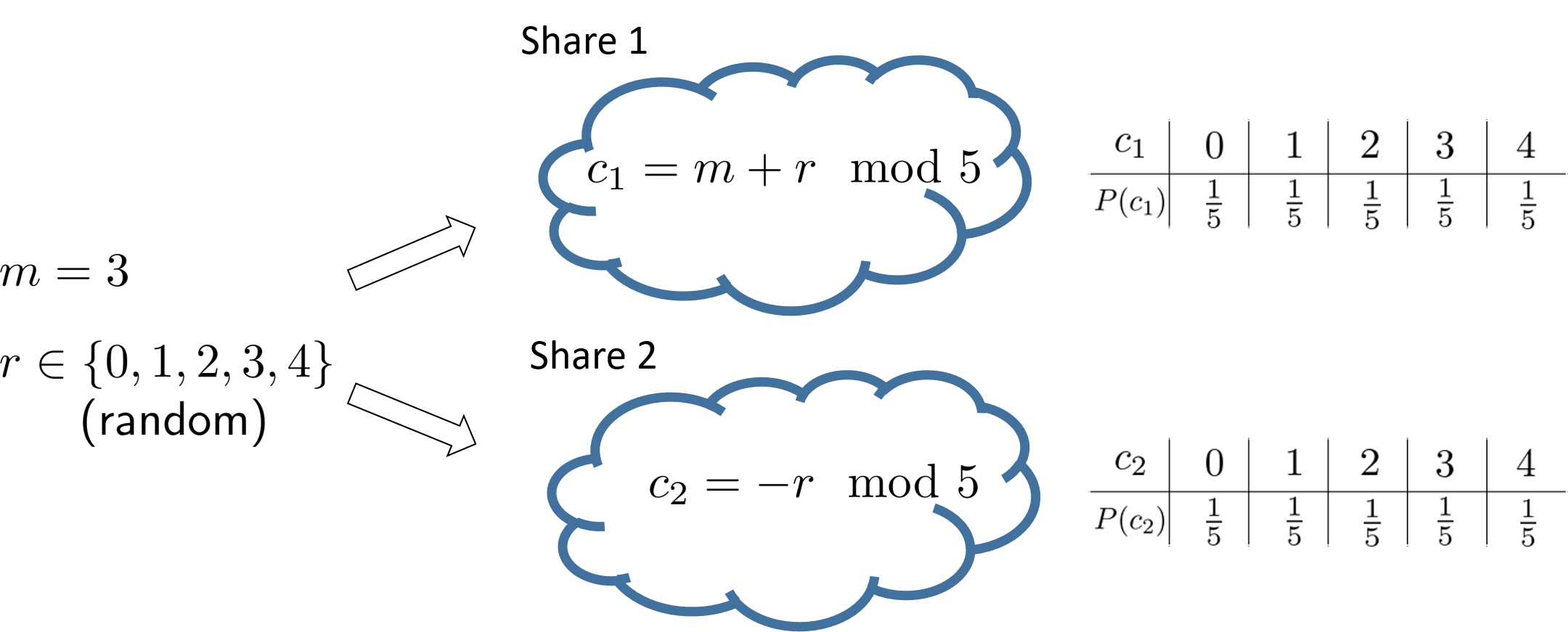}
		\caption{
			A basic example of secret sharing.
		}
		\label{fig:secretsharing}
	\end{figure}

	The most basic secret sharing with two computing parties, supporting additions, is described as follows.
	Let $m\in\Z_q$ be a message that should be kept securely.
	A share for the first party is generated as $c_1 := m + r \bmod q$, where $r\in\Z_q$ is a random number sampled from the uniform distribution over the set $\Z_q$.
	And, the share for the other party is given by $c_2 := -r \bmod q$, so that $m = c_1 + c_2 \bmod q$.
	By doing so, the distribution of each share on each party follows the uniform distribution over the set $\Z_q$, which means that referring to only one share does not give any information about the message $m$.
	We note that a new random number must be sampled whenever a message is split to shares.
	See Figure~\ref{fig:secretsharing} for an example with $\Z_q=\Z_5 = \{0,1,2,3,4\}$ and $m=3$.

	The method itself is quite simple, but this scheme as a cryptosystem is homomorphic with respect to addition and integer multiplication;
	let two messages $m\in \Z_q$ and $m'\in\Z_q$ be split into two shares, as $m= c_1+c_2 \!\!\mod q$ and $m' = c_1'+c_2'\!\!\mod q$, respectively.
	And, let the first shares $c_1\in\Z_q$ and $c_1'\in\Z_q$ be kept by the first operation unit and let the other shares $c_1'\in\Z_q$ and $c_2'\in\Z_q$ be kept by the second operation unit.
	Then, the addition operation ($m+m'\!\!\mod q$) can be performed by the operation units in parallel, as
	\begin{align*}
		\text{Operation unit 1}:\qquad c_1^{\sf add} &= c_1 + c_1'\mod q\\
		\text{Operation unit 2}:\qquad c_2^{\sf add} &= c_2 + c_2'\mod q
	\end{align*}
	where the operation units cannot learn any information of $m$, $m'$, or $m+m'\!\!\mod q$.
	When the outcomes $c_1^{\sf add}$ and $c_2^{\sf add}$ as shares are collected from both the units, then the plaintext outcome can be restored, because $(m+m'\!\!\mod q) = (c_1^{\sf add} + c_2^{\sf add}\!\!\mod q)$.
	And, for a vector $\vec m \in\Z_q^n$ of $n$-messages split as $\vec m = \vec c_1 + \vec c_2~\mathrm{mod}\,q$ by generating $n$-random numbers in $\Z_q$,
	multiplication by (un-encrypted) integer matrices can also be done by the operation units in parallel;
	for a matrix $K\in\Z^{m\times n}$,
	let the operation units compute
	\begin{align*}
		\text{Operation unit 1}:\qquad  c_1^{\sf mult} &= K\cdot \vec c_1 \mod q\\
		\text{Operation unit 2}:\qquad  c_2^{\sf mult} &= K\cdot \vec c_2 \mod q.
	\end{align*}
	Then, it is easy to check that $(K\cdot \vec m \!\!\mod q) = (c_1^{\sf mult} + c_2^{\sf mult}\!\!\mod q)$.
	For general schemes, one may refer to Shamir's secret sharing \cite{shamir79}, and many extended results and methods can be found in literature \cite{karnin83,brickell89,beimel11}.
	
	\begin{figure}[b]
		\centering
		\includegraphics[width=.6\textwidth]{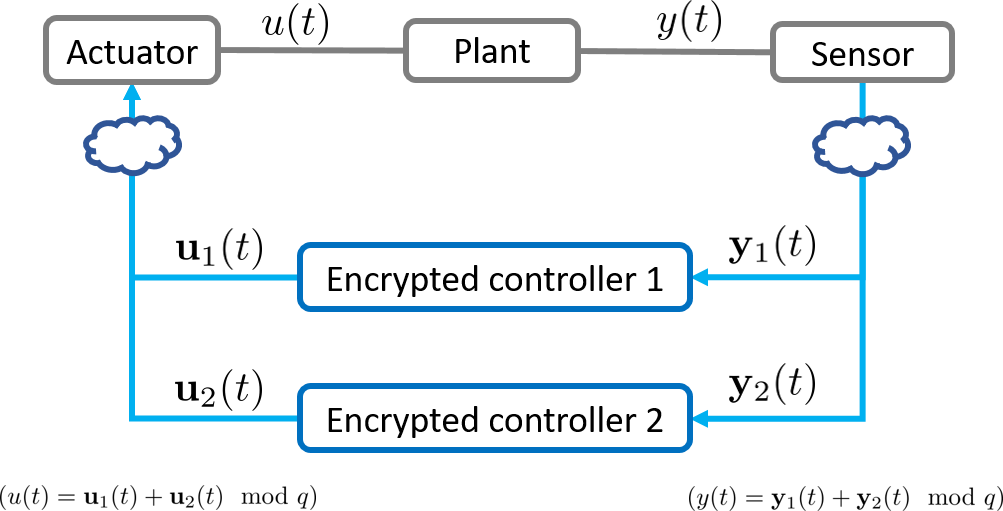}
		\caption{
			Configuration of encrypted controller based on secret sharing with two operation units.
		}
		\label{fig:secretsharing_based}
	\end{figure}

	We have seen that secret sharing enables offloading linear operations without disclosing information, given that the computing parties do not collude with others.
	Contrary to homomorphic encryptions demanding a relatively large amount of storage and computational resources on the operation device, secret sharing based methods can be a good alternative for applications where computational resources are limited but multiple operation units are available.
	For example, it is proposed in \cite{darup19cdc} that the use of secret sharing for encrypted control can reduce the time consumption significantly, per each unit of operation (compared to homomorphic encryption).
	Figure~\ref{fig:secretsharing_based} describes a linear controller based on secret sharing; the sensor measurement $y(t)$ is distributed, as shares $y_1(t)$ and $y_2(t)$, to two controllers, and each controller performs the linear operation and sends the outcome to the actuator, respectively.

	\begin{figure}[t]
		\centering
		\begin{subfigmatrix}{4}
			\subfigure[]{\includegraphics[width=.48\textwidth]{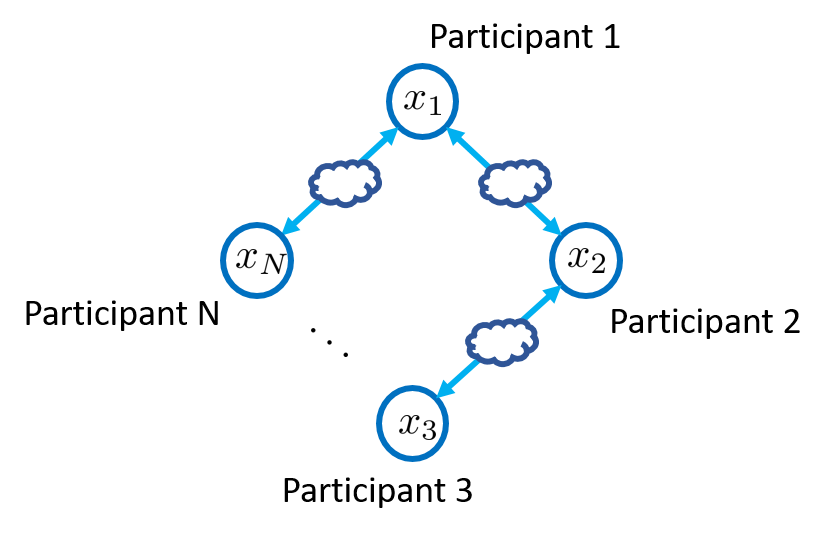}\label{fig:state_decomposition1}}
			\subfigure[]{\includegraphics[width=.48\textwidth,keepaspectratio]{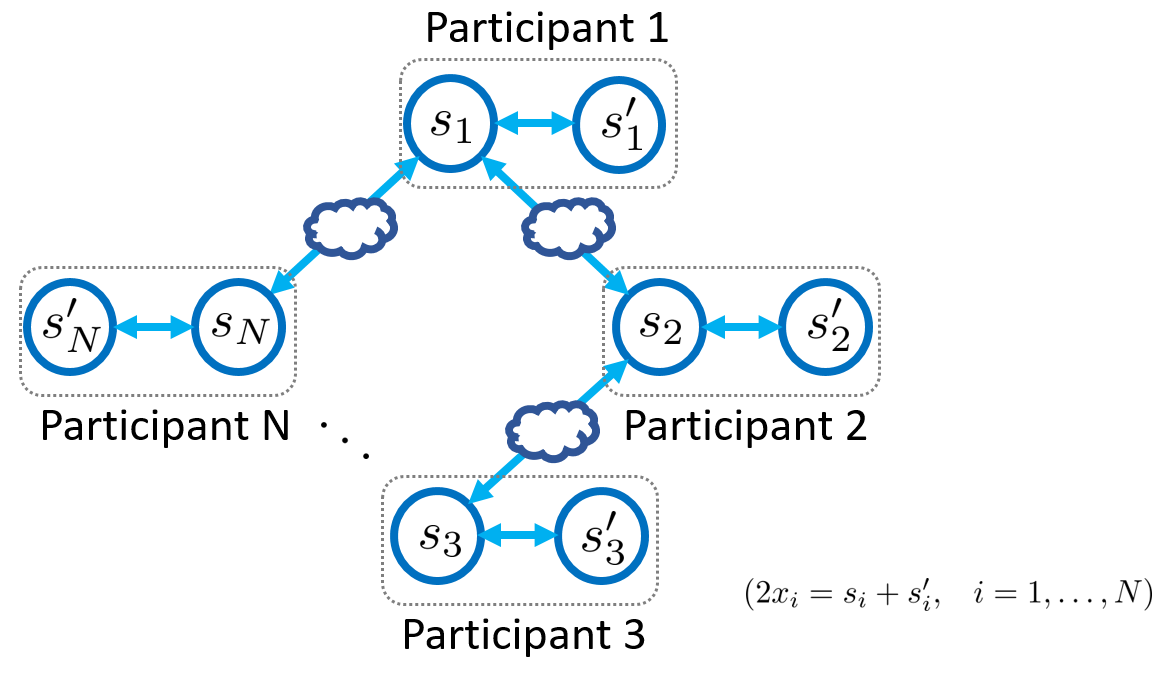}\label{fig:state_decomposition2}}
		\end{subfigmatrix}
		\caption{(a) Conventional distributed protocol. (b) Distributed protocol based on secret sharing.
		}
		\label{fig:state_decomposition}
	\end{figure}
	
	Secret sharing has also been employed for dealing with privacy problems in distributed computation protocols, where multiple participants  collaboratively evaluate a common function of interest.
	For example, average consensus can be considered as in Figure~\ref{fig:state_decomposition1}, where the participants compute the average of the individual values by communicating with their neighbors. 
	In terms of privacy, the goal is to perform the computation while each individual value is not  disclosed to other parties.
	Methods based on differential privacy have been frequently employed as in \cite{mo16tac,nozari17aut}, where a trade-off between privacy and performance is inevitable, depending on the size of injected noises.
	Then, the secret sharing has been introduced for distributed systems; for example, Figure~\ref{fig:state_decomposition2} describes a method proposed in \cite{wang19tac} for privacy-preserving average consensus.
	By splitting each individual value to two shares, transmitting only one of them to the neighbors directly, and reflecting the other share to the protocol ``indirectly,'' the distributed protocol can yield the same outcome without sacrificing the privacy of each individual value.
	An advantage compared to differential privacy is that there is no performance penalty for keeping the data private.

	{\it Operation unit and attack model:}
	Secret sharing based control systems introduce multiple operation units which perform the computation on the distributed shares in parallel, and thus they consider an adversarial model compromising some portion of the units.
	Specifically, ``$(t,N)$-threshold scheme'' denotes that an adversary needs to get $t$-shares out of the $N$-shares (distributed to the units) to recover the message. For instance, the example described in Figure~\ref{fig:secretsharing} can be seen as (2,2)-threshold secret sharing.
	
	{\it  Challenges:}
	Whereas secret sharing allows for efficient computation between the shares of messages,	the issue of possible collusion between the units that reveals the value of the message is a main issue in this framework.
	Furthermore, additional potential risk in practice would be that, if there is an external party (not participating in the computation) who succeeds in collecting the operation units’ information, then the private data can also be exposed.
	Another challenge is that, analogous to methods solely based on homomorphic encryption, general nonlinear functions over secret shares may be hard to implement and currently they are limited.
	There is a trade-off between further operation abilities on the shares and sacrifice of the security, so that it can be studied and improved in the future;
	to perform operations other than addition, 	interactions between the computing units or relaxation of the security model is required.
	For example, in \cite{schlor21cdc}, computation of polynomials of degree $N$ results in a $(2,N)$-threshold secret sharing, where the private message is revealed if any two of the shares are collected.

	There are several works exploiting secret sharing to enhance the privacy of control systems, where many of them also employ homomorphic encryption or other cryptographic schemes, together with use of communication between the operation units.
	We refer to them as ``multi-party computation'' based schemes, which will be discussed in the next subsection.
	In general, secret sharing is often regarded as a sort of multi-party computation scheme, as it also utilizes multiple parties for the computation. Nonetheless, for the sake of detailed comparison and discussion in Section~\ref{subsec:discussion1}, we distinguish the secret sharing algorithms from the multi-party computation; secret sharing desirably does not utilize communication between the parties during the computation, whereas for the multi-party computation, the parties essentially communicate with other parties and collaborate for each unit of computation.

	\subsection{Multi-party computation based control}\label{subsec:mpc}
	
	Finally, we introduce approaches based on multi-party computation.
	The terminology multi-party computation is usually used broadly and often includes the homomorphic encryption and secret sharing based methods discussed in the previous subsections.
	Nonetheless, by ``multi-party computation,'' throughout the article, we mean a narrower framework employing homomorphic encryption, secret sharing, and/or other cryptographic primitives, where the parties cooperate for each unit of computation, by several interactions and communications between them.
	
	\begin{figure}[h]
		\centering
		\begin{subfigmatrix}{4}
			\subfigure[]{\includegraphics[width=.35\textwidth]{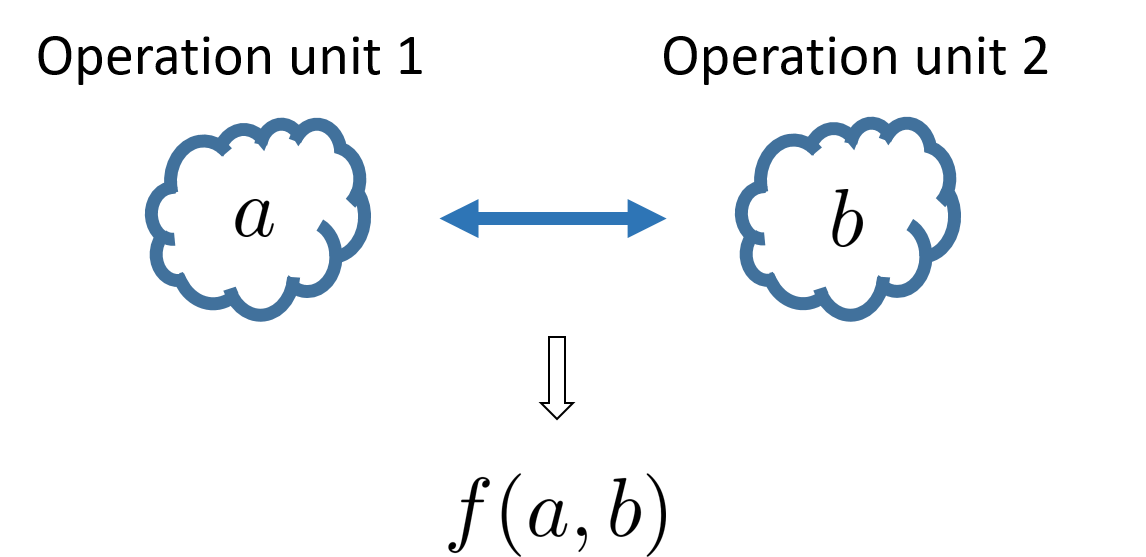}\label{fig:mpc1}}
			\subfigure[]{\includegraphics[width=.35\textwidth,keepaspectratio]{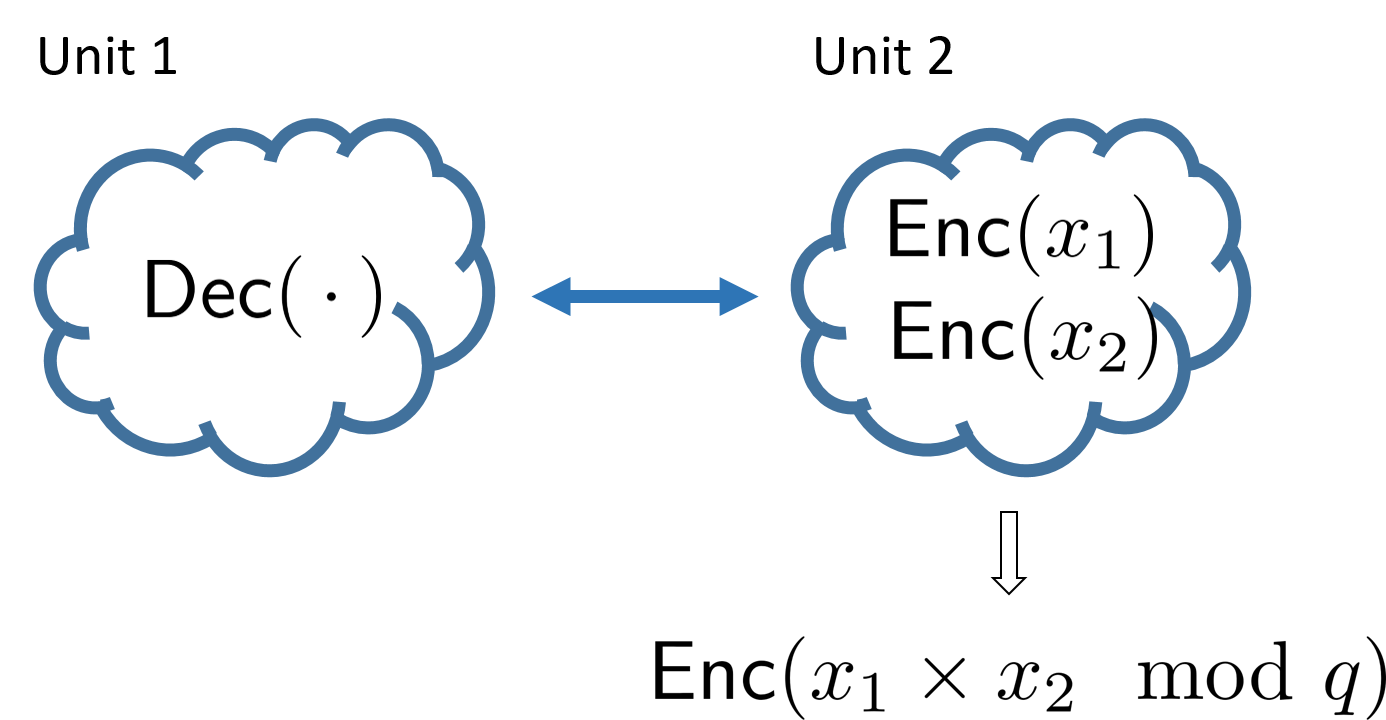}\label{fig:mpc2}}
		\end{subfigmatrix}
		\caption{(a) Configuration of two-party computation. (b) Encrypted multiplication based on two-party computation.
		}
		\label{fig:mpc}
	\end{figure}
	
	The scheme is to let multiple participants collaboratively evaluate a function, where the inputs are given from the participants, but the privacy of each input should be kept.
	That is, each participant’s input should be used for the joint computation, but it should not be learned by any other participant.\footnote{The goal can be defined formally with a simulation paradigm which comprises broader security guarantees such as detecting malicious behaviors with data privacy (we refer to \cite{lindell2017simulate}). In this article, we rather focus on the data privacy aspects.} 
	See Figure~\ref{fig:mpc1} describing a configuration of two-party computation as a basic case; the two units jointly compute the value of $f(a, b)$ while the arguments $a$ and $b$ are not informed by the other unit.

	Figure~\ref{fig:mpc2} shows a basic example of encrypted multiplication by two-party computation.
	Let Unit 2 have two encrypted numbers, Unit 1 have the decryption key, and let the encryption $\Enc$ be additively homomorphic but not multiplicatively homomorphic. That is, Unit 2 can perform encrypted addition using the function $\ast$ as in \eqref{eq:additive}, but it cannot do the multiplication for the two encrypted messages $\Enc(x_1)$ and $\Enc(x_2)$ directly, by itself.
	Then, the objective is to let Unit 2 obtain the multiplication outcome of two messages as encrypted,
	with help of Unit 1,
	while Unit 1 should not learn the values of $x_1$ and $x_2$, and Unit 2 should not obtain Unit 1's decryption key.

\begin{figure}[h]
	\centering
	\includegraphics[width=.8\textwidth]{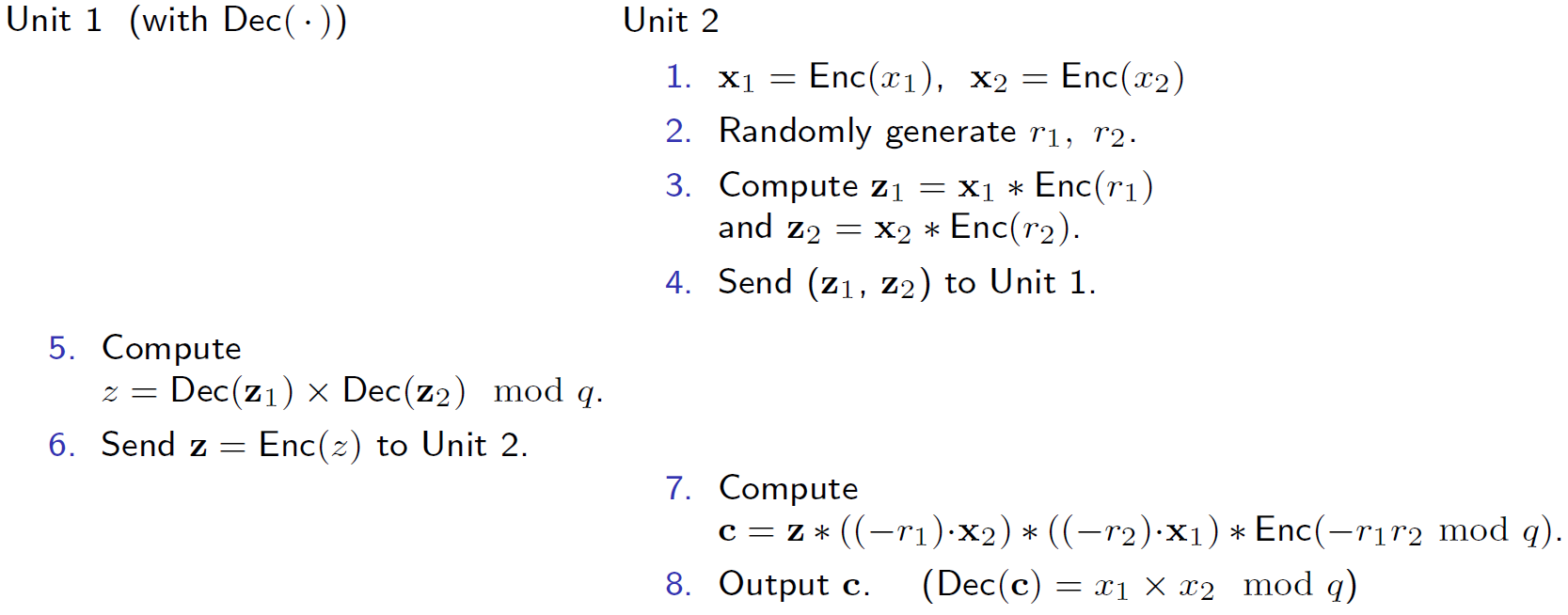}
	\caption{
		A basic method for two-party encrypted multiplication.
	}
	\label{fig:mpc_algorithm}
\end{figure}
	
	Then, Figure~\ref{fig:mpc_algorithm} shows how the two-party computation for encrypted multiplication is performed. Essentially, Unit 2 will send the two messages to Unit 1, and Unit 1 will do the multiplication after decrypting the received ciphertexts. Unit 1 should not know the message values, so Unit 2 generates two random numbers $r_1$ and $r_2$ and add to $\xx_1$ and $\xx_2$ using the additively homomorphic property, and then transmit the ``masked'' messages. Then, even though Unit 1 decrypts the messages, it cannot learn anything from the values $(x_1+r_1\!\!\mod q)$ and $(x_2+r_2\!\!\mod q)$. Nevertheless, now Unit 1 can multiply the decrypted messages,	encrypt the outcome again, and send it back to Unit 2. Finally, the transmitted ciphertext $\zz$ contains the message as $(x_1+r_1)(x_2+r_2)\!\!\mod q$, and since Unit 2 has the information of $r_1$, $r_2$, $\Enc(x_1)$, and $\Enc(x_2)$, it can cancel out the terms $r_1x_2$, $r_2x_1$, and $r_1r_2$, over encrypted data. As a consequence, Unit 2 obtains the outcome whose decryption is equal to $x_1\times x_2\!\!\mod q$. It is also clear that Unit 2 does not learn any additional information because it only computes over encrypted messages during the process.

	We have seen that two-party computation enables encrypted multiplication, despite that the ability of multiplication is not given from the homomorphic cryptosystem. In general, a main advantage of multi-party computation is that {\it it can allow various functions other than addition and multiplication while exploiting additively homomorphic encryption only, by use of multiple parties and communication between them}. Many results have contributed for enabling various functions \cite{damgaard06,nishide07,bogetoft09,dahl12,cramer15}, such as division by integers, inversion of encrypted matrices, and comparison or maximum operations. It is also notable that usually multi-party computation does not make use of	bootstrapping of fully homomorphic encryption so that the required amount of computational resources is not exhaustively large, but it rather exploits communication resources several times for each unit of arithmetic function.

	Meanwhile, similarly to the framework of secret sharing, a crucial assumption on multi-party computation is that the parties do not collude with each other. For example, supposing that Units 1 and 2 in the algorithm in Fig.~\ref{fig:mpc_algorithm} are ``outsourced computers'' performing the encrypted computation on client's private data $x_1$ and $x_2$, the information will be exposed once the units collude and Unit 1 decrypts Unit 2's encrypted data.

\begin{figure}[b]
		\centering
		\includegraphics[width=.6\textwidth]{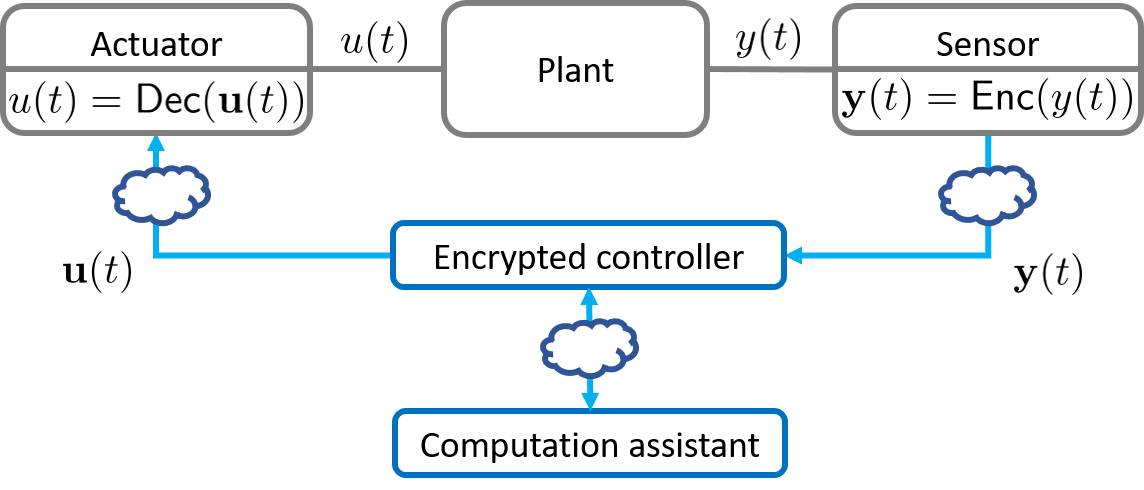}
		\caption{
			Configuration of encrypted controller based on two-party computation.
		}
		\label{fig:multiparty_based}
	\end{figure}

	A basic configuration of the two-party computation based control scheme is described in Figure~\ref{fig:multiparty_based}.
	It considers two collaborative operation units consisting of an ``encrypted controller'' who stores encrypted control data and performs arithmetic supported from the employed homomorphic encryption,
	and a ``computation assistant'' who helps advanced operations that cannot be done with the homomorphic property of cryptosystem solely, by use of two-party computation techniques.
	In case the computation assistant is supposed to have the decryption key of the cryptosystem,
	the actuator can take the role, assuming that the communication between the actuator and the encrypted controller can be bi-directional (for this case, the structure of system becomes similar with that of Figure~\ref{fig:dynamic}, where additional communication is used for ``re-encrypting'' the state).

	Multi-party computation based control employs multiple operation units, which are often regarded as ``out-sourced'' computers, who performs the computation on behalf of the controller designer.
	Thanks to its utility enabling further advanced arithmetic functions, it has been applied to nonlinear controllers which need more than addition and multiplication for operating.
	For example, it has been used for encrypted implementations of ``implicit'' model predictive controllers where projection operation is needed for constrained optimizations \cite{alexandru18cdc,alexandru2020springer}, or extended Kalman filters where matrix inversions and comparison operations are required \cite{gonzalez14}.

	{\it Attack model:}
	Since more than one parties are involved, the adversarial model for multi-party computation framework is similar to that of secret sharing. It can be described similarly according to the number of parties that the adversary is required to compromise to recover the underlying messages. 
	Usually, the security model also considers the behavior that the adversary can carry on; ``honest-but-curious'' model assumes that the adversary follows the protocol correctly (but tries to infer useful information from the observation) whereas ``malicious'' model assumes that the adversary may not follow the protocol correctly and perform an arbitrary behavior (for this case, the protocol aims to enable other participants to detect such behaviors).

	{\it Challenges:}
	Compared to fully homomorphic encryption (which achieves a similar functionality), multi-party computation methods in general require much less computational resources but exploit more communication costs for each unit of arithmetic functions.
	Interestingly, similarly to (fully) homomorphic encryption, evaluating non-arithmetic functions such as division, transcendental functions, and conditional expressions may require much more communications (and computations) than arithmetic functions.
	While there are many specialized multi-party protocols for evaluating a unit of function, it often requires combining them together to design an encrypted system.
	We also remark that constructing a system based on multi-party computation protocol usually involves more complex security analysis and proofs, even in the same honest-but curious model, than that based on (fully) homomorphic encryption, due to the presence of multiple parties and interactions between them where the adversary tries to compromise.

	\subsection{Discussion and comparison}\label{subsec:discussion1}

	Finally, Table~\ref{tab} puts all the discussed approaches and their features together, and compares them with each other, where general trade-offs between the encrypted control approaches are found.
	To this end, for simplicity, let us abbreviate homomorphic encryption as ``HE,'' fully homomorphic encryption as ``FHE,'' and secret sharing as ``SS,'' in this subsection.
	Multi-party computation protocols, utilizing homomorphic encryption, are particularly denoted by ``MPC-HE,'' so that they are distinguished from secret sharing.
	
	\begin{table}[h]
		\caption{Trade-offs between cryptographic tools (compared to HE).}
		\label{tab}
		\begin{center}
			\begin{tabular}{|c|c|c|c|c|}
				\hline
				&$\begin{matrix}
					\textsf{HE}
				\end{matrix}$ &
				$\begin{matrix}
					\textsf{FHE}
				\end{matrix}$
				& {\sf SS} & {\sf MPC-HE}
				\\
				\hline
				\hline
				{\sf feature}
				& - & {\colr\sf computation-intensive}& $\begin{matrix}
					\textsf{\colg operation without}\\\textsf{\colg homomorphic encryption}
				\end{matrix}$ & {\colr\sf communication-intensive}\\
				\hline
				{\sf operation unit}
				& {\sf single} & {\sf single} &{\sf\colo multiple} & {\colo\sf multiple}\\
				\hline
				{\sf strengths}&-
				&$\begin{matrix}
					\textsf{\colg no limitation}\\\textsf{\colg on operations}
				\end{matrix}$ & {\colg\sf computationally efficient} & $\begin{matrix}
					\textsf{\colg advanced operations}\\\textsf{\colg without bootstrapping}
				\end{matrix}$ \\
				\hline
				{\sf weaknesses} & $\begin{matrix}
					\textsf{limitation on}\\\textsf{operations}
				\end{matrix}$
				&$\begin{matrix}
					\textsf{\colr high computational}\\\textsf{\colr complexity}
				\end{matrix}$ & $\begin{matrix}
					\textsf{limitation on operations,}\\\textsf{\colo collusion issues}
				\end{matrix}$& $\begin{matrix}
					\textsf{\colr communication delay,}\\\textsf{\colo collusion issues}
				\end{matrix}$ \\
				\hline
			\end{tabular}		
		\end{center}
	\end{table}
	
	One natural observation is that ``the more resources are utilized, then the more operation abilities (over encrypted data) the scheme obtains''.
	Comparing HE and FHE, it can be seen that FHE utilizes much more computational resources, but it guarantees that any sort of operation circuit can be implemented.
	Comparing HE and MPC-HE, on the other hand, it can be understood that additional use of communication resources for MPC-HE schemes mitigates the limited operation issue of HE;
	e.g., the ability of HE is limited to addition and multiplication over integers,
	but MPC-HE can implement many more advanced functions.
	
	Another aspect in comparison is that ``use of multiple operation units'' brings improvement of computational efficiency, but it also brings possible security issues together.
	Comparing HE and MPC-HE again, MPC-HE takes advantage of multiple parties so that it obtains more operation abilities,
	and comparing HE and SS,
	SS also introduces multiple operation units so that it improves computational efficiency in the sense that it keeps the ability of linear operation still amenable, despite that it does not use homomorphic encryption.
	However, a main opportunity cost for utilizing multiple operation units is possible security issues;
	it should be guaranteed that the participating units never collude with the others,
	and it might also be vulnerable to eavesdropping from external adversaries.
	
	\begin{figure}[h]
		\centering
		\includegraphics[width=.38\textwidth,keepaspectratio]{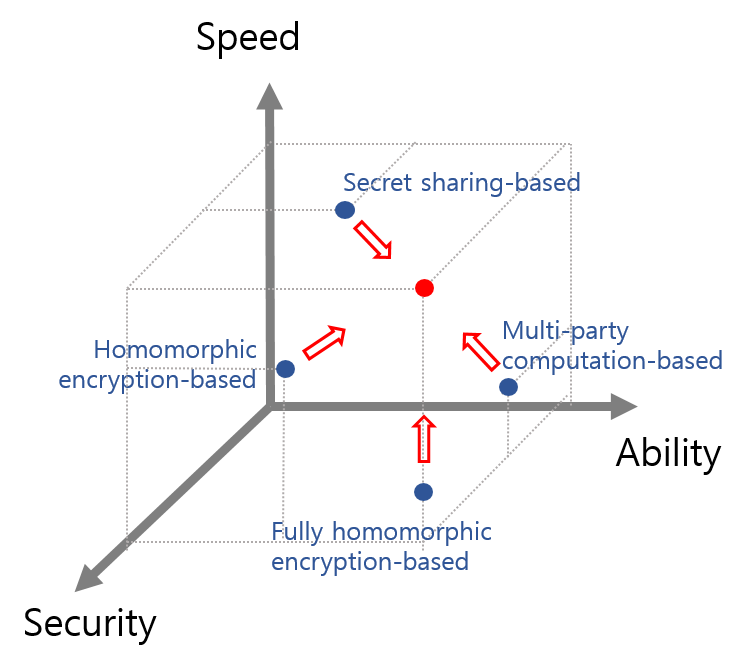}
		\caption{Comparison of encrypted control approaches and future research directions.
		}
		\label{fig:space}
	\end{figure}
	
	A similar observation can be made in a different way.
	Let Figure~\ref{fig:space} compare the four approaches, in terms of speed (operation amount per unit of time), ability (enabled operations over encrypted data), and security (mainly considers issues of collusion and external eavesdropping).
	In terms of ``speed,'' SS may be a good option if the system can be implemented using multiple computers.
	Regarding ``ability,'' if the encrypted controller design includes complicated nonlinear functions, either FHE based method or MPC-HE based protocol can be considered, with intensive use of computational or communication resources, respectively.
	And, for applications in which use of multiple operation units is not allowed because of security matters, use of HE or FHE would be appropriate.
	In short, each encrypted control application should consider its required sort of operations, available resources, and appropriate security model, for choosing its approach.

	Lastly, Figure~\ref{fig:space} also considers future research directions for the approaches, 	which would be to deal with and improve on their respective weaknesses.
	HE based schemes would aim for overcoming the constraint of limited operation, and proposing practical bootstrapping techniques would be of main interest for FHE based methods.
	For the sake of control application, SS and MPC-HE based schemes should resolve possible security issues due to collusion or external eavesdropping.
	At the same time, developing nonlinear functions using SS and reducing the communication overhead with MPC-HE would be of challenges, respectively.

\section{Introduction to linear dynamic systems using LWE-based homomorphic encryption}\label{sec:dynamic}

So far we have discussed broad approaches on encrypted control and current challenges.
To handle those issues, new problems that have not been considered in the control community are often formulated.
In this section, let us be more focused on the approach based on homomorphic encryption only, and introduce a specific problem and current solutions, on implementation of dynamic systems over encrypted data.

Implementing an encrypted controller is to convert a ``given'' controller over plaintext (described in Figure~\ref{fig:config1}) to a system over ciphertext (as Figure~\ref{fig:config2}).
Let a linear dynamic controller has been designed and given as
\begin{align}\label{eq:controller_given}
\begin{split}
x(t+1) &= Fx(t) + Gy(t)\\
u(t) &= Hx(t) + J y(t)\\
x(0) &= x_0
\end{split}
\end{align}
where $x(t)\in\R^\ell$ is the state, $y(t)\in\R^\ppp$ is the sensor measurement transmitted from the plant (input of the controller), $u(t)\in\R^\mmm$ is the actuation input fed back to the plant (output of the controller), and the matrices $\{F,G,H,J\}$ and the initial state $x_0\in\R^\ell$ are given as controller parameters.
Throughout the paper, we assume that the trajectories of the signals $\{x(t),y(t),u(t)\}_{t=0}^{\infty}$ are bounded.

Then, the implementation of the controller \eqref{eq:controller_given} over encrypted data can be divided into two steps.
\begin{enumerate}
\item {\it Conversion of given controller to operate over integers.}
As discussed in Section~\ref{subsec:HE}, most homomorphic cryptosystems allow no more than addition and multiplication over encrypted integers, unless bootstrapping (of fully homomorphic encryption) is used.
Thus, the given system \eqref{eq:controller_given} with the parameters $\{F,G,H,J,x_0\}$, which are generally non-integers,
should be converted to operate over integers using addition and multiplication only.
The input-output relation (performance) of the re-constructed controller should be equivalent to the given controller.

\item {\it Choice of homomorphic cryptosystem and application to dynamic operation}.
The employed cryptosystem should be appropriate to the dynamic operation of the controller.
If the implemented controller is to perform recursive multiplication by an encrypted parameter for each time iteration (just as in \eqref{eq:controller_given}), the cryptosystem should be homomorphic with respect to such recursive operation.
To that end, it can be expected that the encryption should be homomorphic with respect to both addition and multiplication, and it should allow recursive multiplication as encrypted, for an unlimited number of times.

\end{enumerate}

Let us begin with the second issue in the next subsection. We introduce a Learning With Errors (LWE) based encryption called ``GSW-LWE'' proposed in \cite{chillotti16}, which can be understood as a combination of ``Gentry-Sahai-Waters (GSW)'' encryption scheme \cite{gentry13} and a basic LWE-based encryption \cite{lindner11}. Detailed explanations on the encryption scheme and algorithms will be followed by a discussion on its benefits to dynamic operation.

\subsection{An LWE-based cryptosystem for encrypted recursive multiplication}\label{subsec:LWE}

We first introduce a basic version of LWE-based cryptosystem.

\subsubsection{Encryption method, security, and additively homomorphic property \cite{regev2009,lindner11}}

Let the set $\Z_q=\{0,1,\ldots,q-1\}$, $q\in\N$, be the space of messages for encryption.
We have seen in Section~\ref{subsec:SS} and Figure~\ref{fig:secretsharing} that ``adding a random number $r\in\Z_q$ and taking modulo operation by $q$'' is an easy way to conceal a message $m\in\Z_q$.
Though it is quite simple, it may be impractical in the sense that the information of the number $r\in\Z_q$ is required for restoring the message, which is newly sampled whenever a message is encrypted.

To be a cryptosystem using the same ``key'' for each encryption, LWE based schemes generate ``random-like'' numbers from a fixed portion of data, called an encryption key; let a row vector $\sk = [\sk_1,\sk_2,\ldots,\sk_n]\in\Z^n$, $n\in\N$, be chosen, which we will call the encryption key or the secret key.
Then, the idea of LWE-based schemes is that the number $c\in\Z_q$, generated as
\begin{align}\label{eq:enc1}
\begin{split}
c&=\begin{bmatrix}
\sk_1&\sk_2&\cdots&\sk_n
\end{bmatrix}\cdot \begin{bmatrix}
a_1\\a_2\\\vdots\\a_n
\end{bmatrix}+e\mod q\\
&=\sk \cdot a + e \mod q,
\end{split}
\end{align}
looks ``almost uniformly random'' in the space $\Z_q$,
where the components of $a=[a_1,a_2,\ldots,a_n]^\top $ are random numbers in $\Z_q$ (i.e., sampled from the uniform distribution over $\Z_q$), and $e\in\Z$ is a ``small error'' sampled from $N(0,\sigma)$, which denotes the zero-mean discrete Gaussian distribution with standard deviation $\sigma>0$.

The claim that the method \eqref{eq:enc1} generates ``almost random'' numbers is in fact exactly what the Learning With Errors (LWE) problem considers.
Let a ``sufficient'' amount of samples $\{c_i\}_{i=1}^{N}$ in  $\Z_q$ be generated using the key $\sk\in\Z^n$, as
\begin{align}\label{eq:LWE}
\begin{split}
c_1 &= \sk \cdot \vec a_1 + e_1 \mod q\\
c_2 &= \sk \cdot \vec a_2 + e_2 \mod q\\
&~\vdots\\
c_N &= \sk \cdot \vec a_N + e_N\mod q,
\end{split}
\end{align}
where the components of $a_i\in\Z_q^n$ are uniformly sampled from $\Z_q$ and $e_i\in\Z$ is sampled from the distribution $N(0,\sigma)$, for each $i=1,\ldots N$.
Then, given the information of $\{c_i\}_{i=1}^N$ and $\{\vec a_i\}_{i=1}^{N}$, the LWE problem is to find out if the data $\{c_i\}_{i=1}^N$ are generated through the process \eqref{eq:LWE}, or are just uniformly generated random numbers.
If this problem is hard to solve, it implies that the information of the secret key $\sk$ is ``secure'' from unauthorized parties receiving or eavesdropping $\{\vec a_i\}_{i=1}^{N}$ and $\{c_i\}_{i=1}^N$ only.
And, another version of LWE problem is to solve the equation \eqref{eq:LWE} and find the value of the key $\sk$, from given data of $\{c_i\}_{i=1}^N$ and $\{\vec a_i\}_{i=1}^{N}$.

Note that the equation \eqref{eq:LWE} is defined over the space $\Z_q$ with modulo operation, so that the problem is different from the simple least square problem over real numbers.
It has been known that the LWE-problems are hard; for example, it has been proven in \cite{regev2009} that
the latter version of problem is hard as ``worst-case lattice problems,'' and it has also been known as ``post-quantum cryptography'' \cite{chen16}.

The level of ``hardness'' is determined by the parameters; the dimension $n$ for the key $\sk\in\Z^n$ and the vector $a\in\Z_q^n$, the modulus $q$, and the standard deviation $\sigma$ for the error distribution $N(0,\sigma)$.
For example, in \cite{lindner11}, it has been suggested to choose the parameters to satisfy
\begin{equation}\label{eq:lambda}
n \log q \ge \frac{\lambda +110}{7.2}\cdot \left(\log\left(\frac{\sqrt{2\pi}\sigma}{q}\right)\right)^2,
\end{equation}
in order for ``$\lambda$-bit security,'' which means more than $2^\lambda$-times of iterations of a certain computation are required to solve the LWE problem.
Typically, to ensure a higher level of security, a larger dimension $n$ is chosen for the key $\sk\in\Z^n$.

\newcommand{\nnnn}{\mathfrak{n}}

Then, thanks to the security guaranteed from the LWE problems, we may define an encrypted algorithm with the key\footnote{Usually, the components of $\sk$ are generated by sampling from the distribution $N(0,\sigma)$, the same distribution for the errors.} $\sk\in\Z^n$, as
\begin{equation}
\text{for}~m\in\Z_q,\quad
\Enc(m) := \begin{bmatrix}
m+ \sk \cdot a+e\\
a
\end{bmatrix}\mod q\quad \in\Z_q^{\nnnn},\qquad \nnnn:= n+1,
\end{equation}
where the components of the column vector $a\in\Z_q^n$ are uniformly sampled from $\Z_q$ and the error $e\in\Z$ is sampled from $N(0,\sigma)$ analogous to \eqref{eq:enc1}.
We omit the argument of the key $\sk$ from all the cryptosystem algorithms, for simplicity.
Note that, the encryption outcome becomes an $(\nnnn=n+1)$-dimensional vector consisting of elements in $\Z_q$; as LWE problems suppose that the vector $a$ be public, so that it can be treated as a part of encrypted message (it will be used for decryption, to restore the message).
It can be understood that every encrypted message consists of  ``message part,'' ``random-like part,'' and ``error part,'' as
\begin{equation}\label{eq:three}
\Enc(m) = \begin{bmatrix}
m\\ 0_{n\times 1}
\end{bmatrix} + \begin{bmatrix}
\sk \cdot a \\ a
\end{bmatrix} + \begin{bmatrix}
e \\ 0_{n \times 1}
\end{bmatrix} \mod q,
\end{equation}
where the information of the message is contained in the first component.

As the encryption is to add a random-like number $\sk\cdot a$ to the message together with an error, it can be decrypted by simply canceling out the same number $\sk \cdot a$; let the decryption for a ciphertext $\cc\in\Z_q^\nnnn$ be defined as
\begin{equation}\label{eq:dec}
\Dec(\cc):= \begin{bmatrix}
1& -\sk
\end{bmatrix}\cdot \cc \mod q\quad \in\Z_q,
\end{equation}
considering that the ``random-like term'' $[\sk\cdot a, a^\top]^\top$ in \eqref{eq:three} as a column vector will be canceled out whenever the secret key $[1,-\sk]$ is multiplied from the left, regardless of the random part $a$.
Indeed, it is obvious that
\begin{equation}\label{eq:encdec}
\Dec(\Enc(m)) = m+e \mod q
\end{equation}
holds, when the ciphertext $c$ is simply a newly encrypted message.
The cryptosystem being described is ``symmetric'' as a basic case, where the same key $\sk$ is used for both the encryption and decryption. For a public-key LWE-based cryptosystem in which encryption is possible without knowledge of the key $\sk$, interested readers may refer to \cite{lindner11}.

It can be seen in \eqref{eq:encdec} that the error $e$, ``injected'' during encryption because of security, may perturb the decryption outcome not equal to the original message.
To negate the effect of errors, a scale factor can be used for the messages; let us neglect the probability that a sample from the distribution $N(0,\sigma)$ is larger than a particular multiple $n_0\sigma$ of $\sigma$, $n_0\in\N$, and assume that every error $e$ sampled from $N(0,\sigma)$ is such that $ |e|\le n_0\sigma$.
Then, by modifying the encryption as
\begin{equation}\label{eq:enc_scaled}
\Enc_\LLL(m):=\Enc(\LLL m~~\mathrm{mod}~ q),
\end{equation}
with some scale factor $\LLL\in\N$ such that $\LLL/2 > n_0\sigma$, the message can be restored without error as
\begin{equation}\label{eq:scale1}
\Dec(\Enc_\LLL(m)) = \LLL m + e\mod q \quad \implies \quad \left\lceil\frac{\LLL m + e\mod q}{\LLL} \right\rfloor =m,
\end{equation}
as long as $0 < m <(q/\LLL) - (1/2)$ so that $0\le \LLL m + e < q$.

As one of the simplest versions of LWE based schemes, it can be easily seen that the encryption $\Enc$ is homomorphic with respect to addition.
Let the addition of two encrypted messages be defined simply by the component-wise modular addition as
$$\cc_1\in\Z_q^\nnnn,~~ \cc_2\in\Z_q^\nnnn\qquad \mapsto \qquad \cc_1+ \cc_2 \mod q \quad \in\Z_q^\nnnn. $$
Then, since the decryption $\Dec$ is nothing but multiplying a row vector, by the distributive law, it is obvious that
\begin{align}\label{eq:add}
\begin{split}
\Dec(\cc_1+ \cc_2 \mod q) &= \begin{bmatrix}
1 & -\sk
\end{bmatrix} \cdot (\cc_1 + \cc_2)\mod q\\
&= \Dec(\cc_1)+ \Dec(\cc_2) \mod q
\end{split}
\end{align}
holds for every ciphertexts $\cc_1\in\Z_q^\nnnn$ and $\cc_2\in\Z_q^\nnnn$.
It means that the modular addition for the messages in $\Z_q$ can be performed as encrypted, by the same (component-wise) modular addition over the ciphertext space $\Z_q^\nnnn$, because the decryption of the computation outcome matches to the addition of the plaintext messages $\Dec(\cc_1)$ and $\Dec(\cc_2)$.

Multiplication of a ciphertext $\cc_1\in\Z_q^\nnnn$ by an un-encrypted integer $k\in\Z$ can be done analogously, as
\begin{equation}\label{eq:constant_LWE}
k\in\Z, ~~ \cc_1\in\Z_q^\nnnn \qquad \mapsto \qquad k\cdot \cc_1 \mod q \quad \in\Z_q^\nnnn,
\end{equation}
which is simply the component-wise modular multiplication by the constant $k$. It is also clear that
\begin{align}\label{eq:const}
\begin{split}
\Dec(k\cdot \cc_1 \mod q ) &= \begin{bmatrix}
1 &-\sk
\end{bmatrix} \cdot (k\cdot \cc_1)\mod q\\
&= k\cdot \Dec(\cc_1) \mod q,
\end{split}
\end{align}
so the decryption outcome of integer multiplication for encrypted messages matches to that over plaintexts.

In general, LWE-based cryptosystems are also multiplicatively homomorphic, which means they allow multiplication by an encrypted number to ciphertexts as well as by the un-encrypted integer multiplication as in \eqref{eq:constant_LWE}. Nonetheless, in fact, there is no ``trivial way'' of performing multiplication of a ciphertext $\cc_1\in\Z_q^\nnnn$ to another ciphertext $\cc_2\in\Z_q^\nnnn$, since the component-wise multiplication over $\Z_q^\nnnn$ would not work (i.e., decryption outcome of component-wise multiplication would not match with multiplication between the messages), and it seems there is no other ``intuitive way'' of multiplying two vectors in $\Z_q^\nnnn$ yielding the same form in $\Z_q^\nnnn$. In this regard, a separate encryption method as well as an encrypted multiplication method proposed in \cite{gentry13} and \cite{chillotti16} is introduced in the next subsection.

\subsubsection{Gentry-Sahai-Waters (GSW) scheme and encrypted recursive multiplication \cite{gentry13,chillotti16}}

An idea for multiplication by an encrypted message is to consider a separate encryption method whose outcome is a matrix in $\Z_q^{\nnnn\times \nnnn}$, so that it can be  multiplied to the ciphertexts of column vectors in $\Z_q^\nnnn$ from the left naturally; consider a ``multiplier'' $k\in\Z_q$ encrypted as
\begin{equation}\label{eq:trial}
k\in\Z_q \qquad \mapsto \qquad \CC =  k\cdot I_\nnnn + \begin{bmatrix}
\sk \cdot a_1 & \sk \cdot a_2 & \cdots & \sk \cdot a_\nnnn\\
a_1& a_2& \cdots & a_\nnnn
\end{bmatrix} + \begin{bmatrix}
e_1 & e_2 & \cdots & e_\nnnn \\
0_{n\times 1}&0_{n\times 1}&\cdots&0_{n\times 1}
\end{bmatrix}\mod q\quad \in\Z_q^{\nnnn\times\nnnn}
\end{equation}
where $a_i\in\Z_q^n$ is a uniformly random vector and $e_i$ is an error sampled from $N(0,\sigma)$ for each $i=1,\ldots \nnnn$, respectively. The massage part $k\cdot I_\nnnn$ in \eqref{eq:trial} is added by a matrix consisting of ``random-like'' vectors plus errors analogous to \eqref{eq:three}, so obviously it is also an LWE-based encryption.

Then, it can be checked if the multiplication outcome of
\begin{equation}\label{eq:trial_mult}
\CC\in\Z_q^{\nnnn\times\nnnn},~~\cc\in\Z_q^\nnnn \qquad\mapsto\qquad \CC\cdot\cc \mod q \quad \in\Z_q^\nnnn
\end{equation}
(multiplication of the matrix $\CC\in\Z_q^{\nnnn\times\nnnn}$ from \eqref{eq:trial} to a ciphertext vector $\cc\in\Z_q^\nnnn$) has the message parts multiplied with each other; let us decrypt the outcome \eqref{eq:trial_mult} by multiplying $[1,-\sk]$ from the left (just as in \eqref{eq:dec}), which yields
\begin{align*}
&\Dec(\CC \cdot \cc \mod q)\\
&\quad =\begin{bmatrix}
1& -\sk
\end{bmatrix} \cdot \left( k\cdot I_\nnnn + \begin{bmatrix}
\sk \cdot a_1  & \cdots & \sk \cdot a_\nnnn\\
a_1&  \cdots & a_\nnnn
\end{bmatrix} + \begin{bmatrix}
e_1 & \cdots & e_\nnnn \\
0_{n\times 1}&\cdots&0_{n\times 1}
\end{bmatrix}   \right)\cdot \cc \mod q\\
&\quad =\begin{bmatrix}
1& -\sk
\end{bmatrix} \cdot \left( k\cdot I_\nnnn +  \begin{bmatrix}
e_1  & \cdots & e_\nnnn \\
0_{n\times 1}&\cdots&0_{n\times 1}
\end{bmatrix}   \right)\cdot \cc \mod q\\
 &\quad= k\cdot \Dec(\cc) + \begin{bmatrix}
 e_1&\cdots& e_\nnnn
 \end{bmatrix}\cdot \cc \mod q.
\end{align*}
It can be seen that the encryption is homomorphic with respect to the multiplication, in the sense that the decryption outcome has the message $\Dec(\cc)$ multiplied by $k$. However, it can be found that the injected errors $[e_1,\ldots e_\nnnn]$, whose components were smaller than $n_0\sigma$, have been multiplied by the ciphertext $\cc\in\Z_q^\nnnn$ consisting of large numbers. Then, the size of the error ``grows'' too large and it will even dominate the message part, so that the computation result will not be correct at all. Thus, the described method cannot be applied as it is, so modified as follows.

Considering that the problem of the method \eqref{eq:trial_mult} was due to the large size of the ``multiplicand'' $\cc\in\Z_q^\nnnn$ amplifying the errors, the method in \cite{gentry13} called ``Gentry-Sahai-Waters (GSW)'' scheme suggests that the multiplication be done in a different manner, to reduce the size of the multiplicand;
consider multiplication of a vector $\xxx\in\Z_q^\nnnn$ by $k\in\Z_q$, represented as
\begin{equation}\label{eq:mult}
k\cdot \xxx = k\cdot \left( \sum_{i=0}^{d-1} \nu^i \cdot \xxx_i \right)
= k\cdot \begin{bmatrix}
I_\nnnn & \nu \cdot I_\nnnn & \cdots & \nu^{d-1} \cdot I_\nnnn
\end{bmatrix}\cdot \begin{bmatrix}
\xxx_0\\\xxx_1\\\vdots\\\xxx_{d-1}
\end{bmatrix},
\end{equation}
where $\nu\in\N$ is a base chosen such that $ \nu^{d-1}< q \le \nu^d$ with some $d\in\N$, so that the vector $\xxx\in\Z_q^\nnnn$ is represented by ``$\nu$-ary numeral system'' as $\xxx=\sum_{i=0}^{d-1} \nu^i \cdot \xxx_i $ with some vectors  $\{\xxx_i\}_{i=0}^{d-1}$ of non-negative integers such that $\|\xxx_i\| < \nu$, $\forall i$. Then, by defining
$$G :=  \begin{bmatrix}
I_\nnnn & \nu \cdot I_\nnnn & \cdots & \nu^{d-1} \cdot I_\nnnn
\end{bmatrix}\qquad \text{and}\qquad D(\xxx):=\begin{bmatrix}
\xxx_0\\\xxx_1\\\vdots\\\xxx_{d-1}
\end{bmatrix}, $$
and rewriting \eqref{eq:mult} as
$$ k\cdot \xxx = (k\cdot G)\cdot D(\xxx), $$
one can have the size of the multiplicand reduced as  $\|D(\xxx)\|< \nu$, thanks to the ``decomposition'' function $D$.
For example,
consider the equation $77=7\times 11 $ re-written as
$$ 77 = \left(7\times  \begin{bmatrix}
	2^0&2^1&2^2&2^3
\end{bmatrix}\right)\times\begin{bmatrix}
1\\1\\0\\1
\end{bmatrix} $$
with $d=4$, $\nu = 2$. It can be seen that the norm of the multiplicand (which was $11$) is reduced.

From this observation, the GSW scheme slightly modifies the encryption algorithm \eqref{eq:trial}, as
$$\Enc'(k):= k\cdot G + \begin{bmatrix}
\sk \cdot a_1 & \sk \cdot a_2 & \cdots & \sk \cdot a_{d\nnnn}\\
a_1& a_2& \cdots & a_{d\nnnn}
\end{bmatrix} + \begin{bmatrix}
e_1 & e_2 & \cdots & e_{d\nnnn} \\
0_{n\times 1}&0_{n\times 1}&\cdots&0_{n\times 1}
\end{bmatrix}\mod q\quad \in\Z_q^{\nnnn\times d\nnnn},$$
by substituting $I_\nnnn$ in \eqref{eq:trial} by the matrix $G$ and increasing the column dimension $d$-times. Now, let the multiplication of a ciphertext $\cc\in\Z_q^\nnnn$ by an encrypted message $\Enc'(k)$, be performed as
\begin{equation}\label{eq:mult_GSW}
\Enc'(k)\in\Z_q^{\nnnn\times d\nnnn},~~\cc\in\Z_q^\nnnn\qquad \mapsto \qquad \Enc'(k) \cdot D(\cc) \mod q\in\Z_q^\nnnn,
\end{equation}
which enlarges the dimension of $\cc$ by the decomposition first and then multiply to the encrypted multiplier. Then, let us decrypt the outcome and check the multiplicatively homomorphic property and the growth of the error again, as
\begin{align*}
&\Dec(\Enc'(k) \cdot D(\cc) \mod q)\\
&\quad =\begin{bmatrix}
1& -\sk
\end{bmatrix} \cdot \left( k\cdot G + \begin{bmatrix}
\sk \cdot a_1  & \cdots & \sk \cdot a_{d\nnnn}\\
a_1&  \cdots & a_{d\nnnn}
\end{bmatrix} + \begin{bmatrix}
e_1 & \cdots & e_{d\nnnn} \\
0_{n\times 1}&\cdots&0_{n\times 1}
\end{bmatrix}   \right)\cdot D(\cc) \mod q\\
&\quad =\begin{bmatrix}
1& -\sk
\end{bmatrix} \cdot \left( k\cdot G +  \begin{bmatrix}
e_1  & \cdots & e_{d\nnnn} \\
0_{n\times 1}&\cdots&0_{n\times 1}
\end{bmatrix}   \right)\cdot D(\cc) \mod q
\\
&\quad= 
\begin{bmatrix}
1& -\sk
\end{bmatrix}\cdot k  \cdot \cc
+ \begin{bmatrix}
 e_1&\cdots& e_{d\nnnn}
 \end{bmatrix}\cdot D(\cc) \mod q
\\
&\quad= k\cdot \Dec(\cc) + \begin{bmatrix}
 e_1&\cdots& e_{d\nnnn}
 \end{bmatrix}\cdot D(\cc) \mod q.
\end{align*}
It turns out that, thanks to the size of the multiplicand $\cc\in\Z_q^\nnnn$ reduced by the decomposition $D$ as $\|D(\cc)\|<\nu$, the effect of the error becomes bounded by a constant proportional to the parameter $\nu$ (recall that the base $\nu$ can be chosen such that $\nu\ll q$). Putting all together, the homomorphic properties of the described cryptosystem is listed as the following proposition.
\begin{prop}\label{prop:GSW}
The following holds.
\begin{enumerate}
\item For every $m\in\Z_q$, it satisfies $\Dec(\Enc(m)) = m+e \mod q$, with some $e\in\Z$ such that $|e|\le n_0\sigma$.\label{prop:1}
\item For any ciphertexts $\cc_1\in\Z_q^\nnnn$ and  $\cc_2\in\Z_q^\nnnn$, they satisfy $\Dec(\cc_1+\cc_2\mod q)= \Dec(\cc_1) + \Dec(\cc_2)\mod q$.\label{prop:2}
\item For any $k\in\Z$ and $\cc\in\Z_q^\nnnn$, they satisfy $\Dec(k\cdot \cc\mod q) = k\cdot \Dec(\cc)\mod q$.
\item For any $k\in\Z_q$ and $\cc\in\Z_q^\nnnn$, they satisfy $\Dec(\Enc'(k)\cdot D(\cc)\mod q)=k\cdot \Dec(\cc)+e \mod q$, with some $e\in\Z$ such that $|e| \le \Delta_\Mult:= d\nnnn\cdot n_0\sigma\cdot \nu$.\label{prop:4}
\end{enumerate}
\end{prop}

The homomorphic properties directly extended to matrix-vector multiplication over encrypted data, by considering component-wise encryption and operations; let us abuse the notation and the algorithms $\Enc$ and $\Dec$ be also applied to vector of messages component-wisely, as
\begin{subequations}\label{eq:component-wise}
\begin{equation}
\Enc\left(
\vec m
\right):= \begin{bmatrix}
\Enc(m_1)\\\Enc(m_2)\\\vdots\\\Enc(m_l)
\end{bmatrix} \in\Z_q^{l\nnnn},
\end{equation}
for $\vec m = \col\{m_i\}_{i=1}^{l}\in\Z_q^l$,
and
\begin{equation}
\Dec(\vec \cc):= \begin{bmatrix}
\Dec(\cc_1)\\\Dec(\cc_2)\\\vdots\\\Dec(\cc_l)
\end{bmatrix}\in\Z_q^l,
\end{equation}
for $\vec \cc = \col\{\cc_i\}_{i=1}^{l}\in\Z_q^{l\nnnn}$, $\cc_i\in\Z_q^\nnnn$, $i=1,\ldots,l$.
And, let matrices consisting of elements in $\Z_q$ be encrypted using the encryption $\Enc'$ component-wisely, as
\begin{equation}
\Enc'(\FFF) = \begin{bmatrix}
\Enc'(\FFF_{11}) & \Enc'(\FFF_{12}) & \cdots & \Enc'(\FFF_{1l_2})\\
\Enc'(\FFF_{21}) & \Enc'(\FFF_{22}) & \cdots & \Enc'(\FFF_{2l_2})\\
\vdots & \vdots & \ddots &  \vdots\\
\Enc'(\FFF_{l_1 1}) & \Enc'(\FFF_{l_1 2})& \cdots & \Enc'(\FFF_{l_1 l_2})
\end{bmatrix}\in\Z_q^{(l_1 \nnnn)\times(d l_2 \nnnn)}
\end{equation}
for $\FFF = [\FFF_{ij}]\in\Z_q^{l_1\times l_2}$, and let the multiplication of a vector of encrypted messages by an encrypted matrix be considered as
\begin{equation}
\Enc'(\FFF)\in \Z_q^{(l_1 \nnnn)\times(d l_2 \nnnn)},~~ \vec \cc\in\Z_q^{l_2\nnnn}\quad\mapsto\quad \Enc'(\FFF)\cdot D(\vec \cc)\mod q,
\end{equation}
where we also let the decomposition $D$ be applied to a vector of encrypted messages component-wisely as $D(\vec \cc):= \col\{D(\cc_i)\}_{i=1}^{l_2}\in\Z_q^{l_2 d\nnnn}$. Then, it is easy to verify that the following proposition holds.
\end{subequations}

\begin{prop}\label{prop:matrix}
For any matrix $\FFF\in\Z_q^{l_1\times l_2}$ and a ciphertext of encrypted messages $\vec \cc\in\Z_q^{l_2\nnnn}$, they satisfy\footnote{Note that $\Delta_\Mult$ is defined from Proposition~\ref{prop:GSW}.\ref{prop:4}.}
\begin{equation}\label{eq:mult_matrix}
\Dec(\Enc'(\FFF)\cdot D(\vec \cc)\mod q) = \FFF\cdot \Dec(\vec \cc)+ \vec e \mod q,
\end{equation}
with some $\vec e\in\Z^{l_1}$ such that $\| \vec e \| \le l_2\Delta_\Mult$.
\end{prop}

Note that the multiplication of an ``LWE type'' ciphertext by a ``GSW type'' encrypted matrix yields an ``LWE type'' ciphertext. The encryption method of $\Enc'$ has been introduced in \cite{gentry13} where multiplication between GSW type ciphertexts has been presented, and the multiplication of an LWE type ciphertext by a GSW type ciphertext has been considered in \cite{chillotti16}.

Now, we further discuss homomorphic property with respect to ``recursive'' operation.
Let us compare the following two types of homomorphic properties:
\begin{align*}
&\text{Property 1}:\quad \Dec(\Enc(\xxx_1)\ast \Enc(\xxx_2)) = \xxx_1\cdot \xxx_2 \mod q,\qquad \forall \xxx_1\in\Z_q,~\xxx_2\in\Z_q\quad (\text{plaintexts})\\
&\text{Property 2}:\quad \Dec(\cc_1\ast\cc_2) = \Dec(\cc_1)\cdot \Dec(\cc_2)\mod q,\qquad \forall \cc_1\in\Z_q^\nnnn,~\cc_2\in\Z_q^\nnnn\quad(\text{ciphertexts})
\end{align*}
where $\ast$ denotes a certain operation over encrypted data that matches to the multiplication over plaintexts.
Each of the properties implies an ability of multiplication over encrypted data, but it can be observed that Property $2$ implies Property $1$, but Property $1$ does not imply Property $2$ conversely, in general; this is because Property $1$ can be applied to ``newly'' encrypted messages only, whereas Property $2$ can be applied to {\it any} ciphertexts.
For example, for a ciphertext $\cc\in\Z_q^\nnnn$ which is a computation outcome as $\cc=\Enc(\xxx_1)\ast \Enc(\xxx_2)\!\!\mod q$ with some $\xxx_1\in\Z_q$ and $\xxx_2\in\Z_q$, it is clear that the property of multiplication is applicable to $\cc$ if the cryptosystem satisfies Property $2$, whereas the multiplication may not be applicable to $\cc$ even if the cryptosystem satisfies Property $1$, because there is no guarantee that there exists a message $\xxx'\in\Z_q$ such that $\cc = \Enc(\xxx')$.

Keeping this observation in mind, let us revisit the property \eqref{eq:mult_matrix} that we obtained.
In the left hand side where the encrypted multiplication is performed, the argument of left multiplier considers GSW type ciphertext $\Enc'(\FFF)$ as newly encrypted message only, but the argument of the right multiplicand considers any LWE type ciphertext, {regardless if is a newly encrypted message or it was used for certain encrypted operations}.
Since the outcome of the multiplication is also an LWE type ciphertext, the property \eqref{eq:mult_matrix} implies that the encrypted multiplication is applicable to LWE type ciphertexts {\it recursively, unlimited number of times}.
Meanwhile, since an error $\vec e$ is added to the encrypted message whenever the multiplication is performed, it is notable that the ``growth of error'' can be accumulated under recursive operation.
In the next subsection, when the LWE based cryptosystem is applied to dynamic control operations, it will be seen that the growth of error accumulated in LWE type ciphertexts under recursive operation can be ``controlled'' under stability.

The effect of error can be reduced by using a scale factor for the LWE type encryption, as seen in \eqref{eq:scale1}; for instance, let the property \eqref{eq:mult_matrix} be rewritten with respect to the encryption \eqref{eq:enc_scaled}, as
$$\Dec(\Enc'(\FFF)\cdot D(\Enc_\LLL(\xxx))\mod q) = \FFF (\LLL\cdot \xxx + \vec e_1)+ \vec e_2 \mod q, $$
where $\xxx\in\Z_q^{l_2}$, the error $\vec e_1\in\Z_q^{l_2}$ from the encryption $\Enc_\LLL$ is such that $\| \vec e_1 \| \le n_0\sigma$, and the error $\vec e_2 \in\Z_q^{l_1}$ from the multiplication is such that $\| \vec e_2\| \le l_2\Delta_\Mult$.
Then, by choosing the parameter $\LLL$ considering the size of grown error such that $\LLL/2 > \|\FFF\|\cdot n_0\sigma + l_2\Delta_\Mult$, the expected outcome $\FFF \cdot \xxx$ can be obtained without the error, as
\begin{equation}\label{eq:scale2}
 \left\lceil\frac{\FFF\cdot (\LLL \xxx +\vec e_1) + \vec e_2\mod q}{\LLL}\right\rfloor = \FFF\cdot \xxx,
\end{equation}
as long as all the components of $\FFF\cdot \xxx$ is less than $(q/\LLL)-(1/2)$ and greater than $0$ so that the modulo operation in \eqref{eq:scale2} does nothing about the argument value. Note that the scale factor $\LLL$ is used for the LWE type encryption only and not used for GSW type encryption, so that the scaled LWE type messages are not multiplied with another scale factor through the multiplication by GSW type encrypted messages.

As a result of the enabled recursive multiplication, a class of dynamic systems, which can be implemented to run over encrypted data exploiting the homomorphic properties, is specified. Consider a system defined over the space $\Z_q$ written as
\begin{align}\label{eq:system_mod}
\begin{split}
\zzz(t+1) &= \FFF \cdot\zzz(t) + \GGG\cdot \yyy(t) \mod q\\
\uuu(t) &= \HHH \cdot\zzz(t) + \JJJ \cdot\yyy(t) \mod q\\
\zzz(0) &= \zzz_0\in\Z_q^\ell,
\end{split}
\end{align}
where $\zzz(t)\in\Z_q^\ell$ is the state with the initial value $\zzz_0$, $\yyy(t)\in\Z_q^\ppp$ is the input, $\uuu(t)\in\Z_q^\mmm$ is the output, and $\{\FFF,\GGG,\HHH,\JJJ\}$ are matrices consisting of elements in $\Z_q$.
Then, an implication of Proposition~\ref{prop:matrix} is that the dynamic operation of \eqref{eq:system_mod} can be performed over encrypted data, as
\begin{align}\label{eq:system_encrypted}
\begin{split}
\zz(t+1) &= \FF\cdot D(\zz(t)) + \GG\cdot D(\Enc(\yyy(t)))\mod q\\
\uu(t) &= \HH\cdot D(\zz(t)) + \JJ\cdot D(\Enc(\yyy(t))) \mod q\\
\zz(0) &= \Enc(\zzz_0),
\end{split}
\end{align}
where $\zz(t)\in\Z_q^{\ell\nnnn}$ is the state with the initial value $\Enc(\zzz_0)$, $\Enc(\yyy(t))\in\Z_q^{\ppp\nnnn}$ is the input, $\uu(t)\in\Z_q^{\mmm\nnnn}$ is the output, and $\{\FF,\GG,\HH,\JJ\}$ are the encryptions of the matrices from \eqref{eq:system_mod}, as $$\FF = \Enc'(\FFF),\quad \GG = \Enc'(\GGG),\quad \HH = \Enc'(\HHH),\quad \JJ = \Enc'(\JJJ). $$
As the configuration of an encrypted system described in Figure~\ref{fig:config2}, it stores the encrypted parameters $\{\FF,\GG,\HH,\JJ,\zz_0\}$, receives the input $\yyy(t)$ as an LWE type newly encrypted signal, and computes the next state $\zz(t+1)$ and the output $\uu(t)$, respectively, using the GSW-LWE matrix multiplication and the addition between LWE type ciphertexts. The performance of the system \eqref{eq:system_encrypted} can be analyzed, by decrypting the signals $\zz(t)$ and $\uu(t)$ and comparing to that of the un-encrypted model \eqref{eq:system_mod}, as the following proposition.

\begin{prop}\label{prop:system_dec}
Consider the messages
$\tilde \zzz(t):= \Dec(\zz(t))\in\Z_q^\ell$ and $\tilde \uuu(t) := \Dec(\uu(t))\in\Z_q^\mmm$ of the encrypted trajectories of \eqref{eq:system_encrypted}.
They obey
\begin{align}\label{eq:system_dec}
\begin{split}
\tilde \zzz(t+1) &= \FFF\cdot \tilde\zzz(t) + \GGG \cdot(\yyy(t) + \Delta_\yyy(t) )+ \Delta_\zzz(t) \mod q\\
\tilde \uuu(t) &= \HHH \cdot\tilde \zzz(t) + \JJJ \cdot(\yyy(t) + \Delta_\yyy(t) ) + \Delta_\uuu(t) \mod q\\
\tilde \zzz(0) &= \zzz_0 + \Delta_0\mod q,
\end{split}
\end{align}
with some
$\Delta_\yyy(t)\in\Z^\ppp$,
$\Delta_\zzz(t)\in\Z^\ell$, $\Delta_\uuu(t)\in\Z^\mmm$, and $\Delta_0\in\Z^\ell$ such that
\begin{equation}\label{eq:error_bound}
	\| \Delta_\yyy(t)\| \le n_0\sigma,\qquad
\| \Delta_\zzz(t)\|\le (\ell + \ppp)\Delta_\Mult,\qquad \|\Delta_\uuu(t) \| \le (\ell + \ppp)\Delta_\Mult,\qquad \|\Delta_0 \| \le n_0\sigma,
\end{equation}
respectively.
\end{prop}

\noindent{\it Proof:}
By Propositions~ \ref{prop:matrix} and \ref{prop:GSW}.\ref{prop:1},
it is obvious that 
$$\begin{bmatrix}
\tilde \zzz(t+1)\\ \tilde \uuu(t)
\end{bmatrix} =\Dec\left( \begin{bmatrix}
\zz(t+1)\\\uu(t)
\end{bmatrix} \right)  = \Dec\left( \begin{bmatrix}
\FF & \GG \\ \HH & \JJ
\end{bmatrix}\cdot D\left(\begin{bmatrix}
\zz(t) \\ \Enc(\yyy(t))
\end{bmatrix}\right) \right) = \begin{bmatrix}
\FFF & \GGG \\ \HHH & \JJJ
\end{bmatrix}\cdot \begin{bmatrix}
\tilde \zzz(t)\\\yyy(t)+\Delta_\yyy(t)
\end{bmatrix} + \begin{bmatrix}
\Delta_\zzz(t)\\\Delta_\uuu(t)
\end{bmatrix} \mod q$$
and
$ \Dec(\zz(0)) = \Dec(\Enc(\zzz_0)) = \zzz_0 + \Delta_0$ hold, with some $\{\Delta_\yyy(t),\Delta_\zzz(t),\Delta_\uuu(t),\Delta_0\}$
satisfying \eqref{eq:error_bound}.
\hfill$\blacksquare$

Proposition~\ref{prop:system_dec} shows that the performance of the encrypted system \eqref{eq:system_encrypted} is equivalent to the system \eqref{eq:system_dec}, which has the same parameters $\{\FFF,\GGG,\HHH,\JJJ,\zzz_0\}$ with that of \eqref{eq:system_mod}.
It is found that the effect of the errors is denoted by  $\{\Delta_\zzz(t),\Delta_\uuu(t),\Delta_0\}$, which can be seen as disturbances or perturbations bounded as \eqref{eq:error_bound}.

In order to have the error effect relatively small, the encryption $\Enc_\LLL$ with the scale factor $\LLL$ can be used for the input $\yyy(t)$ and the state $\zzz(t)$, as
\begin{align}\label{eq:system_encrypted_scale}
\begin{split}
\zz(t+1) &= \FF\cdot D(\zz(t)) + \GG\cdot D(\Enc_\LLL(\yyy(t)))\mod q\\
\uu(t) &= \HH\cdot D(\zz(t)) + \JJ\cdot D(\Enc_\LLL(\yyy(t))) \mod q\\
\zz(0) &= \Enc_\LLL(\zzz_0).
\end{split}
\end{align}
In the following corollary, it can be observed that only the messages in the dynamics are scaled by the factor $\LLL$, while the bound for the errors remains the same.
\begin{cor}\label{cor:scale}
The messages $\tilde \zzz(t) = \Dec(\zz(t))$ and $\tilde \uuu(t)=\Dec(\uu(t))$ of the system \eqref{eq:system_encrypted_scale} obey
\begin{align}\label{eq:system_dec_scale}
\begin{split}
\tilde \zzz(t+1) &= \FFF\cdot \tilde\zzz(t) + \GGG \cdot(\LLL\cdot \yyy(t)+\Delta_\yyy(t)) + \Delta_\zzz(t) \mod q\\
\tilde \uuu(t) &= \HHH \cdot\tilde \zzz(t) + \JJJ \cdot(\LLL\cdot \yyy(t)+\Delta_\yyy(t)) + \Delta_\uuu(t) \mod q\\
\tilde \zzz(0) &=\LLL\cdot  \zzz_0 + \Delta_0\mod q,
\end{split}
\end{align}
where the errors
$\Delta_\yyy(t)\in\Z^\ppp$,
$\Delta_\zzz(t)\in\Z^\ell$, $\Delta_\uuu(t)\in\Z^\mmm$, and $\Delta_0\in\Z^\ell$ satisfy \eqref{eq:error_bound}.
\end{cor}

We defer the discussion about the error effect in the closed-loop system of the plant and the encrypted controller to the next subsection.
It will be seen that if the closed-loop system is stable with respect to ``perturbations,'' then the effect of the errors can be made arbitrarily small by an appropriate choice of the parameter $\LLL$.

In the remaining, we discuss the strengths and weaknesses of the GSW-LWE encryption scheme.
Compared to other LWE-based cryptosystems, an opportunity cost of using the GSW scheme is that it may require a large amount of storage; for example, if the parameters $(q,\sigma_0,d)$ are chosen as $(q,\sigma_0,d)=(2^{48},10,3)$ and the desired bit-security level is $\lambda = {80}$, then the estimation \eqref{eq:lambda} suggest that the dimension $n$ should be chosen larger than $n=7.16\times 10^2$.
It means that a scalar message encrypted by GSW scheme turns into ($d\nnnn^2\approx 1.54\times 10^6$)-dimensional $48$-bit numbers.
Instead, besides the abilities of addition, multiplication, and recursive operation, another benefit of GSW-LWE scheme is that its implementation is simple and easy; in contrast to the cryptosystems requiring the modulus $q$ chosen as a large odd number, there is no such restriction on the choice of the modulus $q$ with the GSW-LWE scheme.
Then, once choosing the modulus $q$ and the base $\nu$ chosen as powers of $2$, it can be found that the encrypted addition and multiplication, and the algorithms $\Enc$ and $\Dec$ can be implemented with simple modular matrix multiplication, where the operations such as $(\cdot \mod q)$ or $D(\cdot)$ can be easily performed using binary bit operations.

An example of computation times consumed for each encrypted operation is shown in Table~\ref{tab:timecheck}.
In spite of a conservative choice of parameters to ensure the security, the operations $\Enc$, $\Dec$, and the multiplication turn out to be relatively fast.
The encryption $\Enc'$ takes a relatively long time, but considering that it will be used for encrypting the control parameters, it can be performed off-line, while the controller is initialized.
The component-wise encryption and operation described in \eqref{eq:component-wise} would require a multiple of operation times as many as the size of the dimension, but a ``packing method'' \cite{genise19} for the GSW-LWE scheme that encrypts a vector or a matrix into a ``single ciphertext'' can be used for reducing the computation time in practice.

\begin{table}[t]
\caption{Computation times for GSW-LWE algorithms for a scalar variable, tested with MATLAB using an Intel i5 processor with $16\mathrm{GB}$ RAM.}
\label{tab:timecheck}
\begin{center}
\begin{tabular}{|c|c|c|}
	\hline
	$(\nnnn,q,d)$ & $(50,2^{48},3)$& $(1000,2^{48},3)$\\\hline
$\Enc/\Dec$ & $0.2\mathrm{ms}$ & 	$2.6\mathrm{ms}$\\\hline
$\Enc'$ & $15\mathrm{ms}$ & $120 \mathrm{ms}$\\\hline
GSW-LWE multiplication & $0.7\mathrm{ms}$ & $19.9\mathrm{ms}$
\\	\hline
\end{tabular}		
\end{center}
\end{table}

\subsection{Conversion of linear controllers to operate over $\Z_q$}\label{subsec:method}

The previous section showed that the dynamic operation of \eqref{eq:system_mod} can be implemented as \eqref{eq:system_encrypted} with all the parameters and the signals encrypted with the GSW-LWE scheme.
Thus, the remaining task for the encrypted dynamic control implementation is to convert the ``given'' dynamic controller \eqref{eq:controller_given} over $\R$ (which is supposed to be designed in advance) to a system of the form \eqref{eq:system_mod} so that it can operate over the space $\Z_q$ based on the modular arithmetic.

Let the system \eqref{eq:system_mod} be rewritten as
\begin{align}
	\begin{split}\label{eq:system_mod_quantized}
\zzz(t+1) &= \FFF \cdot\zzz(t) + \GGG\cdot Q(y(t)) \mod q
\\
\uuu(t) &= \HHH \cdot\zzz(t) + \JJJ \cdot Q(y(t)) \mod q
\\
\zzz(0) &= \zzz_0\in\Z_q^\ell,
	\end{split}
\end{align}
where $Q:\R^\ppp \ra \Z_q^\ppp$ is a quantization function so that the input $y(t)\in\R^\ppp$ of the given controller \eqref{eq:controller_given} can be regarded as the input of the system \eqref{eq:system_mod_quantized} as well.
Now, given the controller \eqref{eq:controller_given}, the problem of interest is to design the parameters $\{\FFF,\GGG,\HHH,\JJJ,\zzz_0\}$ of \eqref{eq:system_mod_quantized}, so that the performance of the system \eqref{eq:system_mod_quantized} over $\Z_q$ is ``practically equivalent'' to that of the controller \eqref{eq:controller_given} over $\R$.
The problem is more specifically stated, as follows.

\begin{prob}\label{prob}
	Given the parameters $\{F,G,H,J,x_0\}$ of the system \eqref{eq:controller_given} consisting of real numbers together with a positive number $\epsilon>0$, find parameters $\{\FFF,\GGG,\HHH,\JJJ,\zzz_0\}$ of integers and a function $Q$ for the system \eqref{eq:system_mod_quantized}, and find a function $g:\Z_q^\mmm \ra \R^\mmm$, such that $\| u(t) - g(\uuu(t))\|\le \epsilon$ for all $t\ge 0$.
\end{prob}

\begin{figure}[t]
	\centering
	\begin{subfigmatrix}{4}
		\subfigure[]{\includegraphics[width=.45\textwidth]{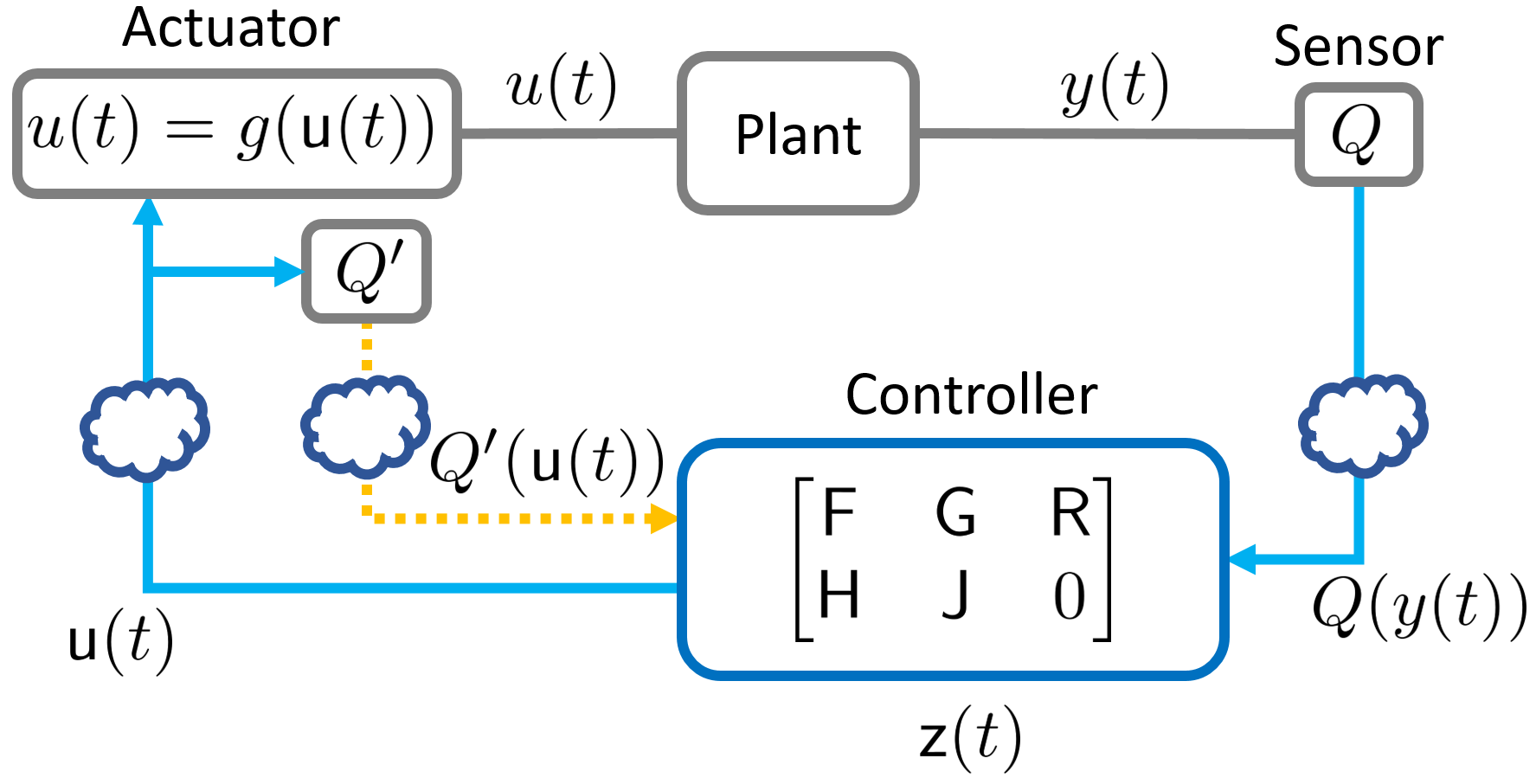}\label{fig:reenc1}}
		\subfigure[]{\includegraphics[width=.45\textwidth,keepaspectratio]{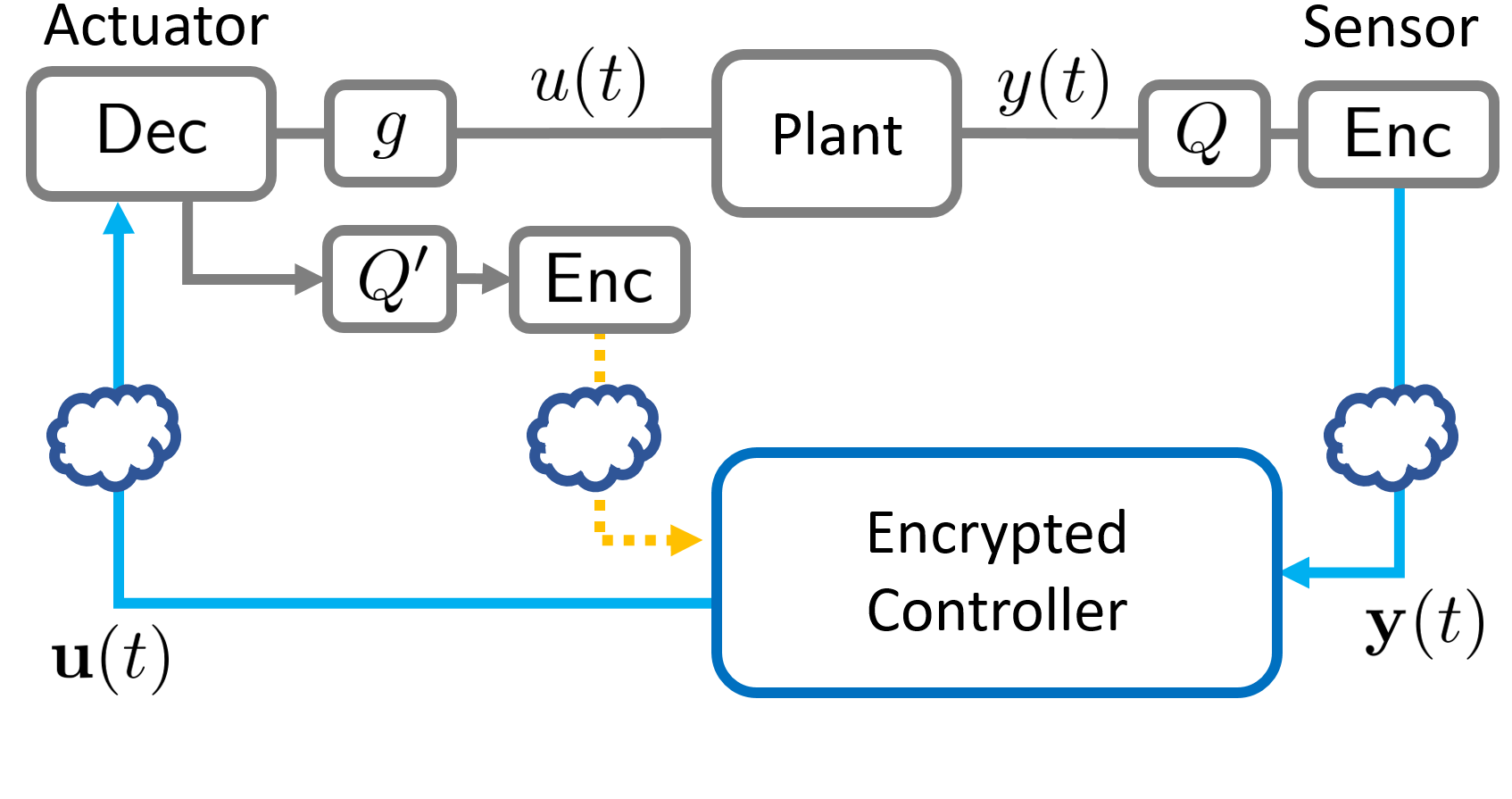}\label{fig:reenc2}}
	\end{subfigmatrix}
	\caption{(a) Block diagram of the system \eqref{eq:system_mod_quantized}. (b) Encrypted implementation of \eqref{eq:system_mod_quantized} with output re-encryption.
	}
	\label{fig:reenc}
\end{figure}

It is desirable that, as in Figure~\ref{fig:config2}, only the encrypted signal of the plant output $y(t)$ is received at the controller, and only the encrypted output of $u(t)$ is transmitted back to the plant.
Nonetheless, from the rationale that the output of the system is supposed to be decrypted at the actuator, an additional assumption can be considered that the decrypted signal of $u(t)$ can be re-encrypted and transmitted back to the controller again; let the system \eqref{eq:system_mod_quantized} be replaced with the following form
\begin{align}\label{eq:system_reenc}
	\begin{split}
		\zzz(t+1) &= \FFF \cdot \zzz(t) + \GGG\cdot Q(y(t)) + \RRR\cdot Q'(\uuu(t))\mod q\\
		\uuu(t) &= \HHH \cdot\zzz(t) + \JJJ \cdot Q(y(t)) \mod q
		\\
		\zzz(0) &= \zzz_0\in\Z_q^\ell
	\end{split}
\end{align}
where a (nonlinear) function $Q': \Z_q^{\mmm}\ra \Z_q^{\mmm}$ can be applied to the decrypted output before it is re-encrypted. See Figure~\ref{fig:reenc} describing the encrypted implementation of \eqref{eq:system_reenc}. Then, a relaxed problem assuming the ``output re-encryption'' can be reformulated as follows.

\begin{prob}\label{prob:reenc}
	Given the parameters $\{F,G,H,J,x_0\}$ of \eqref{eq:controller_given} and $\epsilon>0$, 	find a set of parameters $\{\FFF,\GGG,\HHH,\JJJ,\zzz_0,\RRR\}$ of integers and functions $\{Q,Q'\}$ for the system \eqref{eq:system_reenc} over $\Z_q$, and a function $g$, such that $\|g(\uuu(t)) -  u(t) \|\le \epsilon$,  $\forall t\ge 0$.
\end{prob}

Regarding the stated problems, we note that simply ``scaling up'' the parameters $\{F,G,H,J,x_0\}$ cannot be a solution.
Consider the following example
\begin{subequations}\label{eq:example}
\begin{align}
	\begin{split}
		{x(t+1)} &= {-0.25\times x(t)} +1\\
		u(t)&=x(t)\\
		x(0)&=1
	\end{split}\qquad
	\ra
	\begin{split}
		u(1) &= {0.75},\\
		u(2) &= {0.8125},\\
		u(3) &= {0.796875},\\
		u(4)&= {0.80078125}.
	\end{split}
\end{align}
To keep the precision of the parameter $0.25$ that is recursively multiplied with the state $x(t)$, let it be scaled by $100$ and stored as an integer and the system is represented as
\begin{align}
	\begin{split}
		{\zzz(t+1)} &= {-25\times \zzz(t)} +100^{t+1}\\
		\uuu(t) &= \zzz(t)\\
		\zzz(0)&=1
	\end{split}\qquad
	\ra
	\begin{split}
		\uuu(1) &= {75},\\
		\uuu(2) &= {8125},\\
		\uuu(3) &= {796875},\\
		\uuu(4)&= {80078125},
	\end{split}
\end{align}
\end{subequations}
so that the output $u(t)$ is recovered as $u(t) = \uuu(t)/100^{t}$.
However, it can be seen that the size $|\uuu(t)|$ of the output increases (exponentially) as time goes by, so that no matter how large the modulus $q$ is chosen for the underlying space $\Z_q$ for the system \eqref{eq:system_mod_quantized}, there will be an overflow and the output will be incorrect, in a finite time.

In general, a major issue related to the problem is {\it recursive multiplication of the state matrix consisting of non-integers}; the state matrix $F$ of the given controller in \eqref{eq:controller_given}, which is recursively multiplied with the state $x(t)$, should not be kept as $\lceil F / \sss\rfloor\in\Z^{\ell\time \ell\times \ell}$ with a scale factor $1/\sss> 1$ to keep the precision of the decimal part.
If so, the state $\zzz(t)$ of the implemented system will be multiplied by the scaled matrix $\lceil F / \sss\rfloor$ together with the factor $1/\sss> 1$, so that the size of $\zzz(t)$ (equivalently, the number of decimal places of the state $x(t)$) will increase as time goes by, even if the norm $\|x(t)\|$ of the state is bounded.

In this regard, from the following subsection, we first convert the state matrix to integers without use of scaling, and then convert the rest part of the system to integers. The methods to be introduced are presented in \cite{kim22TAC}, which will be seen as a solution to Problem~\ref{prob:reenc}.

\subsubsection{Conversion of state matrix to integers}

A simple way of changing the parameters of a linear system while keeping the same input-output relation is the ``similarity transformation''.
With an invertible matrix $T\in\Z^{\ell\times \ell}$, let the given controller \eqref{eq:controller_given} be transformed as
\begin{align}\label{eq:system_transformed}
	\begin{split}
		z(t+1) &= TFT^{-1}z(t) + TGy(t)\\
		u(t) &= HT^{-1}z(t) + J y(t)\\
		z(0) &= Tx_0
	\end{split}
\end{align}
which has the same input-output relation with the relation $z(t)=Tx(t)$.
Then, to convert the given state matrix $F$ to integers, one may try to find a transformation matrix $T$ such that the state matrix $F\in\R^{\ell\times\ell}$ consisting of non-integers is turned into integers; i.e., $TFT^{-1}\in\Z^{\ell\times\ell}$.

However, this attempt will not be successful in general, due to the invariant characteristics of the linear systems.
If there exists a transformation $T$ such that $TFT^{-1}\in\Z^{\ell\times\ell}$, then it is obvious that the characteristic polynomial $\det (s I_{\ell}-TFT^{-1})$ of the matrix $TFT^{-1}$ has the coefficients as integers.
Thus, from the invariance that $\det (s I_{\ell}-TFT^{-1})=\det(sI_{\ell} - F)$ no matter what the transform $T$ is chosen, there does not exist a transformation $T$ such that $TFT^{-1}\in\Z^{\ell\times\ell}$ as long as the given state matrix $F$ is such that its characteristic polynomial $\det(sI_{\ell} - F)$ does not have integer coefficients.

Then, considering that the objective of Problem~\ref{prob:reenc} is to convert the given controller into the form \eqref{eq:system_reenc} where the output of the system can be treated as an ``auxiliary input'' at the same time, we first change the state matrix using the auxiliary input and then try to transform it to integers; let us observe that the system \eqref{eq:controller_given} can be rewritten as
\begin{align}\label{eq:auxiliary}
	\begin{split}
	x(t+1)&= (F-RH)x(t) + (G-RJ)y(t)+ Ru(t)\\
	u(t) &= Hx(t) + Jy(t)\\
	x(0) &= x_0
		\end{split}
\end{align}
which has the same relation from the input $y(t)$ to the output $u(t)$, regardless of the choice of the matrix $R\in\R^{\ell\times \mmm}$.
Now, we consider the similarity transformation for the system \eqref{eq:auxiliary}, as
\begin{subequations}\label{eq:auxiliary_transformed}
\begin{align}
		z(t+1)&= T(F-RH)T^{-1}z(t) + T(G-RJ)y(t)+ TRu(t)\label{eq:auxiliary_transformed_state}
		\\
		u(t) &= HT^{-1}z(t) + Jy(t)\label{eq:auxiliary_transformed_output}\\
		z(0) &= Tx_0 \label{eq:auxiliary_transformed_initial}
\end{align}
\end{subequations}
and try to find a transformation matrix $T$ such that the ``converted'' state matrix consists of integers; that is, $$T(F-RH)T^{-1}\in\Z^{\ell\times\ell}.$$

In what follows, it is proposed that the system of the form \eqref{eq:auxiliary} can always be converted to have the state matrix as integers.
First, without loss of generality, we can assume that the pair $(F,H)$ of the given controller \eqref{eq:controller_given} is observable, i.e.,
$$ \mathrm{rank}\left(\begin{bmatrix}
	H\\HF\\\vdots\\HF^{\ell-1}
\end{bmatrix}\right)=\ell.$$
If it is not, ``Kalman observable decomposition'' can be considered so that the pair $(F,H)$ can be reduced to an observable pair; an invertible matrix $W = [W_1^\top,W_2^\top]^\top\in\R^{\ell\times\ell}$ can always be found such that, with $z_1(t) = W_1 x(t)\in\R^{\ell'}$ and $z_2(t)=W_2x(t)\in\R^{\ell-\ell'}$ with some $\ell'\le \ell$, the controller \eqref{eq:controller_given} can be transformed into the form
\begin{subequations}
\begin{align}
	z_1(t+1) &= F_{11} z_1(t) + W_1Gy(t)\label{eq:kalman_obs} \\
	z_2(t+1) &= F_{21}z_1(t) + F_{22} z_2(t) + W_2 Gy(t)\label{eq:kalman_unobs}\\
	u(t) &= H_1 z_1(t) + Jy(t)\label{eq:kalman_output}
\end{align}
\end{subequations}
with
$$WFW^{-1} = \begin{bmatrix}
	F_{11} & 0_{\ell' \times(\ell -\ell')}\\
	F_{21} & F_{22}
\end{bmatrix} \quad \text{and}\quad HW^{-1} = \begin{bmatrix}
H_1 & 0_{\mmm\times(\ell - \ell')}
\end{bmatrix},$$
where the reduced pair $(F_{11},H_1)$ becomes observable.
Then, since the sub-state $z_2(t)$ of the unobservable part does not affect the value of the output $u(t)$ in \eqref{eq:kalman_output}, it suffices to reduce the given controller \eqref{eq:controller_given} to the subsystem \eqref{eq:kalman_obs} with \eqref{eq:kalman_output} only, which is observable and has the same input-output relation.

Now, based on the observability of the pair $(F,H)$, it is proposed that, the matrices $R$ and $T$ can always be found such that the converted system \eqref{eq:auxiliary_transformed} has the state matrix as integers; let the matrix $R$ be found such that the eigenvalues $\{\lambda_i\}_{i=1}^{\ell}$ of the matrix $F-RH$ are distinct integers, for example.
Then, by choosing the transformation matrix $T$ that diagonalizes the matrix $F-RH$, we obtain
$$T(F-RH)T^{-1} = \begin{bmatrix}
	\lambda_1& 0& \cdots& 0\\
	0 &\lambda_2 & \cdots& 0\\
	\vdots &  \vdots & \ddots & \vdots\\
	0&0 &\cdots & \lambda_\ell
\end{bmatrix}\in\Z^{\ell\times\ell}, $$
which is clearly a matrix consisting of integers.
As a result, we have the following lemma.
\begin{lem}\label{lem:1}
	For any observable pair $F\in\R^{\ell\times\ell}$ and $H\in\R^{\mmm\times\ell}$, there exist matrices $R\in\R^{\ell\times\mmm}$ and $T\in\R^{\ell\times\ell}$ such that $T(F-RH)T^{-1}\in\Z^{\ell\times\ell}$.
\end{lem}

If the controller \eqref{eq:controller_given} is an observable single-output system, the observable canonical form of the system can be simply used for the conversion.
Indeed, let $H\in\R^{1\times\ell}$, and let the transform matrix $T$ be found such that the pair $(F,H)$ is transformed into the form
\begin{equation*}
	TFT^{-1} =  \begin{bmatrix}
			0 &0& \cdots & 0 & a_1\\
			1 &0& \cdots & 0 & a_2\\
			0&1&\cdots&0&a_3\\
			\vdots & \vdots& \ddots & \vdots & \vdots\\
			0 &0& \cdots & 1 & a_{\ell}
	\end{bmatrix}\in\R^{\ell\times\ell},\qquad HT^{-1}=\begin{bmatrix}
		0&\cdots&0&1
	\end{bmatrix}\in\R^{1\times\ell}.
\end{equation*}
Then obviously, any matrix $R\in\R^{\ell\times1}$, given by
\begin{equation*}
R =T^{-1}\begin{bmatrix}
	a_1 - k_1\\
	a_2 - k_2\\
	\vdots \\
	a_{\ell} - k_{\ell}
\end{bmatrix},\qquad k_i\in\Z,\quad i=0,\ldots, \ell,
\end{equation*}
will yield the converted state matrix as integers, as
\begin{equation*}
T(F-RH)T^{-1}  = \begin{bmatrix}
	0 &0& \cdots & 0 & k_1\\
	1 &0& \cdots & 0 & k_2\\
	0&1&\cdots&0&k_3\\
	\vdots & \vdots& \ddots & \vdots & \vdots\\
	0 &0& \cdots & 1 & k_{\ell}
\end{bmatrix} \in\Z^{\ell\times\ell}. 
\end{equation*}

\subsubsection{Conversion to system over $\Z$}

So far, we have converted the given controller \eqref{eq:controller_given} to the form \eqref{eq:auxiliary_transformed}, where the matrices $T$ and $R$ are found such that the state matrix $T(F-RH)T^{-1}$ consists of integers.
Now, recalling that the overflow problem described in the example \eqref{eq:example} was due to the state matrix $F$ consisting of non-integers and due to the recursive multiplication of the state matrix $\lceil F/\sss \rfloor\in\Z^{\ell\times\ell }$ scaled by $1/\sss > 1$, it can be expected that such problem would not be found because now the converted system \eqref{eq:auxiliary_transformed} will keep the state matrix $T(F-RH)T^{-1}$ as integers without a scaling factor (that is, $1/\sss=1$).
In this subsection, we show how to implement the system having the state matrix as integers, to operate over the space $\Z$ using addition and multiplication.

The task is to convert all the signals and the parameters of the system \eqref{eq:auxiliary_transformed} to integers using scale factors, except the state matrix which is already of integers.
Note that, any matrix $G\in\R^{\ell\times\ppp}$ (or any vector or scalar) can be stored as integers with arbitrary precision, as $\lceil G/\sss \rfloor\in\Z^{\ell\times \ppp}$, $1/\sss>0$, because the error due to rounding can be arbitrary small as the scale factor $1/\sss$ tends to infinity (as $\sss$ tends to zero), as
\begin{equation}\label{eq:observe_error}
	\left\| G - \sss\left\lceil\frac{G}{\sss} \right\rfloor \right\| \le \frac{\ppp}{2}\cdot \sss.
\end{equation}
Thus, to keep the performance of the system \eqref{eq:auxiliary_transformed} while scaling up the non-integer numbers therein, let $1/\rrr>0$,  $1/\sss>0$, and $\LLL\in\N$ be scale factors, and let the factor $\LLL/(\rrr\sss)$ be multiplied to the both sides of the equations \eqref{eq:auxiliary_transformed_state} and \eqref{eq:auxiliary_transformed_initial}, and the factor $\LLL/(\rrr\sss^2)$ be multiplied to the both sides of the equation \eqref{eq:auxiliary_transformed_output}, as
\begin{align}\label{eq:scale_up}
		\begin{split}
	\frac{\LLL\cdot z(t+1)}{\rrr\sss}&= T(F-RH)T^{-1}\cdot\frac{\LLL\cdot z(t)}{\rrr\sss} + \frac{T(G-RJ)}{\sss}\cdot\frac{\LLL\cdot y(t)}{\rrr}+ \frac{TR}{\sss}\cdot\frac{\LLL\cdot u(t)}{\rrr}
	\\
	\frac{\LLL\cdot u(t)}{\rrr\sss^2} &= \frac{HT^{-1}}{\sss}\cdot\frac{\LLL\cdot z(t)}{\rrr\sss} + \frac{J}{\sss^2}\cdot\frac{\LLL\cdot y(t)}{\rrr}\\
	\frac{\LLL\cdot z(0)}{\rrr\sss} &= \LLL\cdot \frac{Tx_0}{\rrr\sss}.
	\end{split}
\end{align}

Next, we take the rounding operation to the parameters and signals so that we consider the following system
\begin{subequations}\label{eq:system_Z}
\begin{align}
		\ol z(t+1)&= \ol F \cdot \ol z(t) + \ol G \cdot \ol y(t) + \ol R\cdot \LLL\cdot  \left\lceil\frac{\sss^2}{\LLL}\cdot \ol u(t)\right\rfloor, \label{eq:system_Z_state}\\
		\ol u(t) &= \ol H \cdot \ol z(t) + \ol J \cdot \ol y(t) \label{eq:system_Z_output}\\
		\ol z(0) &= \ol z_0,\notag
\end{align}
\end{subequations}
with
\begin{align}\label{eq:rounding_errors}
	\begin{split}
	\ol F &:= T(F-RH)T^{-1},\quad \ol G := \left\lceil \frac{T(G-RJ)}{\sss}\right\rfloor,\quad \ol R := \left\lceil\frac{TR}{\sss}\right\rfloor,\\
	\ol H &:= \left\lceil\frac{HT^{-1}}{\sss}\right\rfloor,\quad \ol J := \left\lceil\frac{J}{\sss^2}\right\rfloor,\quad \ol y(t) :=\LLL\cdot  \left\lceil\frac{y(t)}{\rrr}\right\rfloor,\quad \ol z_0 := \LLL\cdot \left\lceil \frac{Tx_0 }{\rrr\sss} \right \rfloor,
\end{split}
\end{align}
where the state $\ol z(t)\in\Z^{\ell}$, the input $\ol y(t)\in\Z^{\ppp}$, and the output $\ol u(t)\in\Z^{\mmm}$ will keep approximate values of the state $\LLL\cdot z(t)/(\rrr\sss)$, the input $\LLL\cdot y(t)/\rrr$, and  the output $\LLL\cdot u(t)/(\rrr\sss^2)$ in \eqref{eq:scale_up}, respectively.
Note that, the constructed system \eqref{eq:system_Z} is defined over integers and operates using only addition and multiplication, except the operation $\LLL\cdot \lceil (\sss^2/\LLL)\cdot(~) \rfloor$  which ``divides'' the value of $\ol u(t)$ by the factor $1/\sss^2$.

Next, we consider the performance of the system  \eqref{eq:system_Z}.
Define $\tilde z(t) := (\rrr\sss/\LLL)\cdot \ol z(t)$ and $\tilde u(t) : = (\rrr\sss^2/\LLL)\cdot \ol u(t)$. Then, the system \eqref{eq:system_Z} can be rewritten as
\begin{align}\label{eq:system_perturbed_z}
		\begin{split}
	\tilde z(t+1)&= T(F-RH)T^{-1}\tilde z(t) + T(G-RJ)y(t)+ TR \tilde u(t)+e_z(t)
	\\
	\tilde u(t) &= HT^{-1}\tilde z(t) + Jy(t)+e_u(t)\\
	\tilde z(0) &= Tx_0+e_{z,0}
		\end{split}
\end{align}
where $e_z(t)\in\R^{\ell}$, $e_u(t)\in\R^{\mmm}$, and $e_{z,0}\in\R^\ell$ denote the error due to the  rounding (quantization), given by
\begin{align}\label{eq:errors}
\begin{split}
	e_z(t) &= \left(T(G-RJ)y(t) - \left(\sss\left\lceil\frac{T(G-RJ)}{\sss} \right\rfloor\right)\cdot\left(\rrr\left\lceil\frac{y(t)}{\rrr} \right\rfloor\right)\right)+\left(TR\tilde u(t) - \left(\sss\left\lceil\frac{TR}{\sss} \right\rfloor\right)\cdot\left(\rrr\left\lceil\frac{\tilde u(t)}{\rrr} \right\rfloor\right)\right)
		\\
	e_u(t) &= \left(HT^{-1}-\sss\left\lceil\frac{HT^{-1}}{\sss}\right\rfloor \right)\cdot \tilde z(t) + \left(Jy(t) - \left(\sss^2\left\lceil\frac{J}{\sss^2} \right\rfloor\right)\cdot\left(\rrr\left\lceil\frac{y(t)}{\rrr} \right\rfloor\right)\right)
		\\
	e_{z,0} &= Tx_0 - \rrr\sss\left\lceil\frac{Tx_0}{\rrr\sss} \right\rfloor.
	\end{split}
\end{align}
We note that the factor $\LLL$ does not affect the values of $\{e_z(t),e_u(t),e_{z,0}\}$, so does nothing about the performance of \eqref{eq:system_perturbed_z}; as an auxiliary scale factor, it will be used in Section~\ref{subsec:error_growth}, to deal with ``error growth'' of the LWE-based cryptosystem.
One may suppose $\LLL=1$ in this subsection for simplicity.

Then, the performance of the implemented system \eqref{eq:system_Z} can be identified with the given controller model \eqref{eq:controller_given} with presence of perturbation; from \eqref{eq:system_perturbed_z}, it can be easily verified that the state $\tilde x(t) := T^{-1}\tilde z(t)$ and the output $\tilde u(t)$ obey
\begin{align}\label{eq:controller_perturbed}
	\begin{split}
		\tilde x(t+1) &= F\tilde x(t) + Gy(t)+e_x(t)\\
		\tilde u(t) &= H\tilde x(t) + J y(t)+e_u(t)\\
		\tilde x(0) &= x_0+e_0,
	\end{split}
\end{align}
where the errors $e_x(t)\in\R^{\ell}$ and $e_0\in\R^{\ell}$ are given by
\begin{equation}\label{eq:error_z_to_x}
	e_x(t) = T^{-1}e_z(t) + TR e_u(t),
	\qquad e_0 = T^{-1} e_{z,0}.
\end{equation}
Compared to the given controller \eqref{eq:controller_given} without considering the errors, note that $\tilde x(t) = x(t)$ and $\tilde u(t) = u(t)$, $\forall t\ge0$, if $e_0=0$, $e_x(t)=0$, and $e_u(t)$, $\forall t\ge 0$.
The result is that the performance of \eqref{eq:system_Z} is equivalent to that of \eqref{eq:controller_perturbed}, with the relation
\begin{equation}\label{eq:relation_Z}
	\tilde x(t) = \frac{\rrr\sss}{\LLL}\cdot  T^{-1}\ol z(t),\qquad \tilde u(t) = \frac{\rrr\sss^2}{\LLL}\cdot \ol u(t).
\end{equation}

In what follows, we discuss the effect of the perturbations $\{e_x(t),e_u(t),e_0\}$ on the performance of the systems \eqref{eq:system_Z} and \eqref{eq:controller_perturbed}.
The claim is that, if the sizes of the errors $\{e_x(t),e_u(t),e_0\}$ can be made arbitrarily small by appropriate choice of the parameters $\{\rrr,\sss\}$, and if the trajectories $\tilde x(t)$ and $\tilde u(t)$ of \eqref{eq:controller_perturbed} are ``stable'' with respect to the perturbations $\{e_x(t),e_u(t),e_0\}$, then the difference of the trajectories $\tilde x(t)$ and $\tilde u(t)$ from those of the ``ideal'' trajectories $x(t)$ and $u(t)$ can be made arbitrarily small by the choice of $\{\rrr,\sss\}$. And, this will ensure that the performance of the system \eqref{eq:system_Z} is ``practically equivalent'' to the given controller \eqref{eq:controller_given}.
In this regard, we make the following assumption.
\begin{asm}\label{def}
	Given the controller \eqref{eq:controller_given},
	the perturbed model \eqref{eq:controller_perturbed} satisfies the following:
	There exists a function $\delta(\epsilon)>0$ such that for any $\epsilon>0$, if
\begin{equation}
	\|e_0\|\le \delta(\epsilon),~~
	\|e_x(t)\|\le \delta(\epsilon),~~\text{and}~~\|e_u(t)\|\le \delta(\epsilon),~~\forall t\ge 0,\label{eq:def1}
\end{equation}
then $\|\tilde x(t) - x(t) \| \le \epsilon$ and $\|\tilde u(t)-u(t)\|\le \epsilon$ hold for all $t\ge 0$.
\end{asm}

Obviously, if the given controller \eqref{eq:controller_given} itself is stable (i.e., the state matrix $F$ is Schur stable), then the condition of Assumption~\ref{def} is implied. But, we note that a system, in which the condition of Assumption~\ref{def} holds, may not be stable by itself (the matrix $F$ may not be Schur stable).
Typically, if the system \eqref{eq:controller_perturbed} is a part of a closed-loop system, and if it is stable with respect to the perturbations $\{e_x(t),e_u(t),e_0\}$, then the condition of Assumption~\ref{def} will hold.
In such cases, rigorously, the input $y(t)$ of the system may also be affected by the perturbations, but we omit such arguments for simplicity (See \cite[Section II.B]{kim22TAC}).

Finally, the following lemma states that the performance of the converted controller \eqref{eq:system_Z} operating over integers is guaranteed, provided that the trajectories of the given controller is stable with respect to the quantization errors $\{e_x(t),e_u(t),e_0\}$.
For a more rigorous proof under a more general setting, see \cite[Proposition 6]{kim22TAC}.

\begin{lem}\label{lem:Z}
	For any $\epsilon>0$,
	under Assumption~\ref{def},
	there exist $\rrr'>0$ and $\sss'>0$ such that for any $\rrr<\rrr'$, $\sss<\sss'$, and $\LLL\in\N$,
	the controller \eqref{eq:system_Z}  implemented over $\Z$ guarantees that
	$\|(\rrr\sss/\LLL)\cdot T^{-1} \ol z(t) - x(t)\|\le \epsilon$ and $\|(\rrr\sss^2/\LLL)\cdot\ol u(t) - u(t)\|\le \epsilon$ hold, $\forall t \ge 0$.
\end{lem}

\noindent{\it Sketch of Proof:}
Recall the relation \eqref{eq:relation_Z} between the system \eqref{eq:system_Z} and \eqref{eq:controller_perturbed},
with the perturbations $\{e_x(t),e_u(t),e_0\}$ determined by \eqref{eq:errors} and \eqref{eq:error_z_to_x}.
Given $\epsilon>0$ and $\delta(\epsilon)$,
we show that
$\rrr'$ and $\sss'$ can be chosen such that $\rrr<\rrr'$ and $\sss<\sss'$ ensure that \eqref{eq:def1} holds.
Observe that, for any signal $y(t)\in\R^{\ppp}$ and $G\in\R^{\ell \times \ppp}$,
some constants $\theta_1$ and $\theta_2$ can be found such that
\begin{align*}
\left\|Gy(t) - \left(\sss\left\lceil \frac{G}{\sss} \right\rfloor\right)\cdot \left(\rrr\left\lceil \frac{y(t)}{\rrr} \right\rfloor\right)\right\| &\le \frac{\|G\|}{2}\cdot\rrr + \frac{\ppp}{2}\cdot\|y(t)\|\cdot\sss + \frac{\ppp}{4}\cdot\rrr\sss\\
&\le \theta_1\cdot\max\{\|y(t)\|,\theta_2\}\cdot\max\{\rrr,\sss,\rrr\sss\}.
\end{align*}
Then, from \eqref{eq:errors} and \eqref{eq:error_z_to_x}, it can be verified that
\begin{align}
\max\{\|e_x(t)\|,\|e_u(t)\|,\|e_0\|\}&\le\theta'_1 \cdot \max\{\|y(t)\|,\|\tilde x(t)\|,\|\tilde u(t)\|,\theta'_2\}\cdot \max\{\rrr,\sss,\rrr\sss,\sss^2,\rrr\sss^2\}\notag\\
&=:	\alpha(\|y(t)\|,\|\tilde x(t)\|,\|\tilde u(t)\|,\rrr,\sss)\label{eq:alpha}
\end{align}
with some constants $\theta_1'$ and $\theta_2'$.
From the boundedness of the signals of \eqref{eq:controller_given},
let $\max\{\|y(t)\|,\|x(t)\|,\|u(t)\|\}\le M$ with some $M>0$.
Then,
since $\rrr=0$ and $\sss=0$ implies $\alpha =0$ in \eqref{eq:alpha},
we can choose $\rrr'>0$ and $\sss'>0$ such that
$$ \alpha(M,M+\epsilon,M+\epsilon,\rrr',\sss')\le \delta(\epsilon). $$
As a consequence,
by Assumption~\ref{def},
it can be proved that any $\rrr<\rrr'$ and $\sss<\sss'$ ensure that
$$\|\tilde x(t) - x(t)\|\le \epsilon,\quad \|\tilde u(t)-u(t) \| \le\epsilon,\quad \|\tilde x(t)\|\le  M+\epsilon,\quad \|\tilde u(t) \| \le M+ \epsilon, $$
for all $t\ge 0$, using mathematical induction. It completes the proof.\hfill$\blacksquare$

\subsubsection{Conversion to system over $\Z_q$}

All the parameters and the signals of the given controller \eqref{eq:controller_given} are now converted to integers, in the controller \eqref{eq:system_Z}.
Recalling the goal of Problem~\ref{prob:reenc}, we map the parameters and signals of \eqref{eq:system_Z} to the set $\Z_q=\{0,1,\ldots,q-1\}$ of messages for encryption, and further convert the system \eqref{eq:system_Z} to the form \eqref{eq:system_reenc}, so that it can be directly encrypted using the result of Proposition~\ref{prop:system_dec}.

First of all, we note that, given any dynamic system of the form
\begin{align}\label{eq:Z}
	\begin{split}
		\ol x(t+1) &= \ol F' \ol x(t) + \ol G' \ol y(t),\\
		\ol u'(t) &= \ol H' \ol x(t) + \ol J' \ol y(t),\\
		\ol x(0) &= \ol x_0,
	\end{split}
\end{align}
where all the signals and parameters are of integers, the conversion to a system over $\Z_q$ itself is simple; just by taking the modulo operation to the parameters, the signals, and the computation outcomes of \eqref{eq:Z}, it can be converted as
\begin{align}\label{eq:Z_q}
	\begin{split}
		\xxx(t+1) &= \FFF' \xxx(t) + \GGG' \yyy(t)\mod q,\\
		\uuu'(t) &= \HHH' \xxx(t) + \JJJ' \yyy(t)\mod q,\\
		\xxx(0) &= \xxx_0	,
	\end{split}
\end{align}
where
\begin{align}\label{eq:Z_q_matrix}
	\begin{split}
\FFF'&= \ol F' \mod q,\quad
\GGG' = \ol G' \mod q,\quad
\HHH' = \ol H'\mod q,\quad
\JJJ' = \ol J'\mod q,\\
\yyy(t)&= \ol y(t)\mod q,\quad
\xxx_0 = \ol x_0 \mod q.
\end{split}
\end{align}
Since the modulo operation is compatible with addition and multiplication, the conversion \eqref{eq:Z_q} ensures that
\begin{equation}\label{eq:Z_to_Z_q}
	\uuu'(t) = \ol u'(t)\mod q,\quad\forall t\ge 0. 
\end{equation}

Conversely, to recover the output $\ol u'(t)$ of arbitrary integers from the ``projected'' output $\uuu'(t)$ consisting of elements of $\Z_q$, the ``size'' $q$ of the space $\Z_q$ should be chosen appropriately.
A sufficient condition is given in the following proposition; the original output $\ol \uuu'(t)$ can be recovered from $\uuu'(t)$ using the biased modulo operation defined in \eqref{eq:biased_modulo}, if the size of the space $\Z_q$ ``covers'' the range of the output $\ol u'(t)$.

\begin{prop}\label{prop:Z_q}
	Consider two systems \eqref{eq:Z} and \eqref{eq:Z_q} with \eqref{eq:Z_q_matrix}.
	Suppose that the output $\ol u'(t)=\col\{\ol u_i'(t)\}_{i=1}^{\mmm'}$, $\mmm'\in\N$, of \eqref{eq:Z} be bounded as
	\begin{equation}\label{eq:prop41}
		\ol u_i^{\min} \le \ol u'_i(t) \le \ol u_i^{\max},\qquad \forall t \ge 0,\quad \forall i=1,\ldots,\mmm',
	\end{equation}
	with some constants
	$\{\ol u_i^{\min}\}_{i=1}^{\mmm'}$ and $\{\ol u_i^{\max}\}_{i=1}^{\mmm'}$.
	If
	\begin{equation}\label{eq:prop42}
		q \ge \ol u_i^{\max} -  \ol u_i^{\min} + 1,\qquad \forall i=1,\ldots,\mmm',
	\end{equation}
	then $\ol u'(t) = \uuu'(t) 
	~~\mathrm{mod}\,
	(q, \col\{\ol u_i^{\min}\}_{i=1}^{\mmm'} )$ holds for all $t\ge 0$.
\end{prop}
{\it Proof:}
Let the $i$-th component of $\uuu'(t) 
~~\mathrm{mod}\,
(q, \col\{\ol u_i^{\min}\}_{i=1}^{\mmm'} )$ be denoted by $\ol u''_i(t)$.
Note that the operations  $\mathrm{mod}\,q$ and $\mathrm{mod}\,
(q, \col\{\ol u_i^{\min}\}_{i=1}^{\mmm'} )$ add some multiples of $q$ to the components only.
Then, from \eqref{eq:Z_to_Z_q}, it is clear that
$\ol u_i'(t) = \ol u_i''(t) + kq$ with some $k\in\Z$.
Now, the definition \eqref{eq:biased_modulo} ensures that $\ol u_i^{\min} \le \ol u_i''(t) <\ol u_i^{\min}+q $, and \eqref{eq:prop41} and \eqref{eq:prop42} ensure that
$\ol u_i^{\min}\le \ol u'_i(t) < \ol u_i^{\min} + q$.
Hence, $k=0$ and $\ol u_i''(t)=\ol u_i'(t)$.
It completes the proof.
\hfill$\blacksquare$

Based on this observation, we convert the system \eqref{eq:system_Z} to the form \eqref{eq:system_reenc} over $\Z_q$, and determine the parameters and functions therein, to find a solution to Problem~\ref{prob:reenc}.
To determine the modulus $q$ first, we calculate the range of the output $\ol u(t)$ of \eqref{eq:system_Z}; from the boundedness of the output $u(t)=\col\{u_i(t)\}_{i=1}^{\mmm}$ of \eqref{eq:controller_given}, let some constants $\{u_{i}^{\min}\}_{i=1}^{\mmm}$ and $\{u_i^{\max}\}_{i=1}^{\mmm}$ be found such that
\begin{equation}\label{eq:bound_u}
u_i^{\min }\le u_i(t) \le u_i^{\max},\qquad\forall i=1,\ldots,\mmm,\quad \forall t\ge 0.
\end{equation}
Then, given $\epsilon>0$ from Problem~\ref{prob:reenc}, the design of \eqref{eq:system_Z} is supposed to ensure that the output $\ol u(t)=\col\{\ol u_i(t)\}_{i=1}^{\mmm}$ is bounded as $\|(\rrr\sss^2/\LLL) \cdot \ol u(t)-u(t)\|\le \epsilon$, which implies
\begin{equation}\label{eq:bound_Z}
	\LLL\cdot \frac{u_i^{\min} - \epsilon}{\rrr\sss^2} \le \ol u_i(t)\le \LLL\cdot \frac{u_i^{\max} + \epsilon}{\rrr\sss^2},\quad \forall i=1,\ldots,\mmm.
\end{equation}

Now, considering the condition \eqref{eq:prop42},
we choose the modulus $q$ such that
\begin{equation}\label{eq:q}
q \ge \max_{i=1,\ldots, \mmm} \left\{\LLL\cdot  \frac{u_i^{\max} - u_i^{\min}+2\epsilon}{\rrr\sss^2}\right\}+1.	
\end{equation}
Then, even if the operation $\mathrm{mod}\,q$ is taken to the output $\ol u(t)$, as $\uuu'(t) = \ol u(t)~\mathrm{mod}\,q$, the value of $\ol u(t)$ can be recovered from $\uuu'(t)$, as
$$\ol u(t) = \uuu'(t) \mod\left(q,\col\left\{
\LLL\cdot \frac{u_i^{\min}-\epsilon}{\rrr\sss^2}
\right\}_{i=1}^{\mmm}\right),$$
and a real-valued signal approximate to the real output $u(t) \approx (\rrr\sss^2/\LLL) \cdot \ol u(t)$ can also be computed from $\uuu'(t)$, with a function $g:\Z_q^\mmm\ra \R^\mmm$ defined as
$$g(\uuu'(t)):= \frac{\rrr\sss^2}{\LLL}\cdot\left(\uuu'(t) \mod\left(q,\col\left\{
\LLL\cdot \frac{u_i^{\min}-\epsilon}{\rrr\sss^2}
\right\}_{i=1}^{\mmm}\right)\right). $$

Finally, we convert the system \eqref{eq:system_Z} to a system over $\Z_q$, by simply taking the modulo operation, as
\begin{subequations}\label{eq:system_Z_q}
\begin{align}
		\zzz(t+1)&= \FFF \cdot \zzz(t) + \GGG \cdot \yyy(t) + \RRR\cdot Q'(\uuu(t)) \mod q \label{eq:system_Z_q_state}\\
		\uuu(t) &= \HHH \cdot \zzz(t) + \JJJ \cdot \yyy(t) \mod q\label{eq:system_Z_q_output}\\
		\zzz(0) &= \zzz_0\notag
\end{align}
\end{subequations}
with 
\begin{align}\label{eq:matrices}
	\begin{split}
		\FFF &= \ol F \mod q,\quad \GGG = \ol G \mod q,\quad \RRR = \ol R \mod q,\quad \yyy(t) = \ol y(t) \mod q,\\
		\HHH &= \ol H \mod q,\quad \JJJ = \ol J \mod q,\quad \zzz_0 = \ol z_0 \mod q,
	\end{split}
\end{align}
where the term $\LLL\cdot\lceil (\sss^2/\LLL)\cdot \ol u(t) \rfloor$ in \eqref{eq:system_Z} is replaced with the term $Q'(\uuu(t))$, with a function $Q':\Z_q^\mmm\ra\Z_q^\mmm$ defined as
\begin{equation}\label{eq:Q'}
	Q'(\uuu(t)):= \LLL\cdot \left\lceil \frac{\sss^2}{\LLL}\cdot  \left(\uuu(t) \mod\left(q,\col\left\{\LLL\cdot 
	\frac{u_i^{\min}-\epsilon}{\rrr\sss^2}
	\right\}_{i=1}^{\mmm}\right)\right) \right\rfloor \mod q.
\end{equation}
Note that, since the operation $\LLL\cdot\lceil (\sss^2/\LLL) \cdot \ol u(t) \rfloor$ for $\ol u(t)$ is not compatible with the modulo operation, in \eqref{eq:Q'}, the value of $\ol u(t)$ is first computed from $\uuu(t)$, and then the operation is applied.
And, note that, with
$$\yyy(t)=Q(y(t)):= \LLL\cdot \left\lceil\frac{y(t)}{\rrr} \right\rfloor ~~\mathrm{mod}~ q, 
$$
the constructed system \eqref{eq:system_Z_q} takes the same form that Problem~\ref{prob:reenc} considers.

As the end result of Section~\ref{subsec:method}, and as a solution to Problem~\ref{prob:reenc}, the following theorem shows that the performance of the implemented system \eqref{eq:system_Z_q}, which operates based on modular arithmetic over the space $\Z_q$, can be arbitrarily close to that of the given system \eqref{eq:controller_given}, by increasing the scale factors $1/\rrr$ and $1/\sss$, regardless of the factor $\LLL$.

\begin{thm}\label{thm:1}
	Given any controller \eqref{eq:controller_given}
	and $\epsilon>0$, under Assumption~\ref{def}, 	there exist $\rrr'>0$ and $\sss'>0$ such that for any $\rrr<\rrr'$, $\sss<\sss'$, and $\LLL\in\N$, 	the controller \eqref{eq:system_Z_q} 	implemented over $\Z_q$ with \eqref{eq:q} guarantees $\|g(\uuu(t))-u(t)\|\le \epsilon$, $\forall t\ge 0$.
\end{thm}
{\it Proof:}
Consider the system \eqref{eq:system_Z} over $\Z$ as an auxiliary system.
Thanks to Lemma~\ref{lem:Z},
we choose $\rrr'$ and $\sss'$ such that any $\rrr<\rrr'$ and $\sss<\sss'$ ensure that $\|(\rrr\sss^2/\LLL) \cdot \ol u(t) - u(t)\| \le \epsilon$, $\forall t\ge0$.
By the boundedness of the output $u(t)$ as
\eqref{eq:bound_u}
with the constants $\{u_i^{\min},u_i^{\max}\}_{i=1}^{\mmm}$,
it follows that \eqref{eq:bound_Z} holds.
Now, it is enough to show that
\begin{equation}\label{eq:thm}
	\ol u(t) =\left(\uuu(t) \mod\left(q,\col\left\{\LLL\cdot 
	\frac{u_i^{\min}-\epsilon}{\rrr\sss^2}
	\right\}_{i=1}^{\mmm}\right)\right)
\end{equation}
for all $t\ge 0$.
Note that $\yyy(t) = \ol y(t)~\mathrm{mod}\,q$, $\forall t\ge 0$, and $\zzz_0 = \ol z_0~\mathrm{mod}\,q$,
so, from
\eqref{eq:system_Z_output} and \eqref{eq:system_Z_q_output}, it follows that $\uuu(0) = \ol u(0)~\mod\,q$.
Now,
suppose that $\uuu(\tau) = \ol u(\tau)~\mod\,q$ for some $\tau\ge 0$.
Then, the condition \eqref{eq:q} ensures that
\eqref{eq:thm} holds for $t=\tau$,
which can be proved analogously to the proof of Proposition~\ref{prop:4}.
This is followed by
$Q'(\uuu(\tau))= \LLL\cdot \lceil (\sss^2/\LLL) \cdot(\ol u(\tau)) \rfloor$,
so that
it can be easily verified that $\zzz(\tau+1) = \ol z(\tau+1)~\mathrm{mod}\,q$,
by comparing
\eqref{eq:system_Z_state} and \eqref{eq:system_Z_q_state}.
And, from \eqref{eq:system_Z_output} and \eqref{eq:system_Z_q_output},
it is followed by
$\uuu(\tau+1) = \ol u(\tau+1)~\mod\,q$. Hence, the proof is completed.\hfill$\blacksquare$

The result of Theorem~\ref{thm:1} can be summarized as follows. First of all, from the example in \eqref{eq:example}, it has been observed that linear systems having the state matrix as non-integers cannot continue the encrypted operation for an infinite time horizon, because of the limitation of homomorphic encryption. Then, based on the assumption that the encrypted controller can receive a re-encrypted signal of its output, Lemma~\ref{lem:1} has converted the state matrix to integers in \eqref{eq:auxiliary_transformed}, while keeping the same relation from the input $y(t)$ to the output $u(t)$.
Next, the quantization parameters $\rrr$ and $\sss$ have been introduced for the rest controller matrices and signals. And, Lemma~\ref{lem:Z} and Theorem~\ref{thm:1} showed that linear systems having the state matrix as integers can be converted to operate over $\Z_q$, based on modular addition and multiplication, so that they can be implemented over encrypted data.
The performance is guaranteed and its error can be made arbitrarily small, as long as the modulus $q$ for the space $\Z_q$ is chosen as \eqref{eq:q}, so that it covers the range of the output.

A follow-up result proposing a solution to Problem~\ref{prob} can be found in \cite{KSSJ21}, where the parameters $\{\FFF,\GGG,\HHH,\JJJ\}$ are found as time-varying matrices consisting of integers.

\subsection{Encrypted  dynamic system and controlled error growth}\label{subsec:error_growth}

Finally, we combine the GSW-LWE cryptosystem (described in Section~\ref{subsec:LWE}) to the implemented controller \eqref{eq:system_Z_q}  over $\Z_q$ (constructed in Section~\ref{subsec:method}), and discuss the result.

Thanks to Proposition~\ref{prop:system_dec} and Theorem~\ref{thm:1}, the result is straightforward; let the matrices of the controller \eqref{eq:system_Z_q} be encrypted using the GSW encryption $\Enc':\Z_q\ra \Z_q^{\nnnn\times d\nnnn}$ component-wisely, as
\begin{align*}
\FF := \Enc'(\FFF),\quad \GG	:= \Enc'(\GGG),\quad
\RR:=\Enc'(\RRR),\quad
\HH:= \Enc'(\HHH),\quad \JJ := \Enc'(\JJJ),
\end{align*}
and let the initial value $\zzz_0\in\Z_q^\ell$ of the controller state and the controller input $y(t)\in\R^\ppp$ be quantized using the function $Q:\R^\ppp\ra\Z_q^\ppp$ and encrypted using the LWE-based encryption $\Enc:\Z_q\ra\Z_q^\nnnn$, component-wisely, as
$$\zz_0 := \Enc(\zzz_0),\qquad \yy(t) :=
\Enc(Q(y(t))). $$
Then, the encrypted controller is constructed as
\begin{align}\label{eq:controller_encrypted}
	\begin{split}
	\zz(t+1) &= \FF\cdot D(\zz(t)) + \GG \cdot D(\yy(t)) + \RR\cdot D(\uu'(t))\mod q\\
\uu(t) &= \HH \cdot D(\zz(t)) + \JJ \cdot D(\yy(t))\mod q\\
\zz(0) &= \zz_0	
\end{split}
\end{align}
where $\zz(t)\in\Z_q^{\ell\nnnn}$ and $\uu(t)\in\Z_q^{\mmm\nnnn}$ are the encrypted state and output, respectively, and $\uu'(t)\in\Z_q^{\mmm\nnnn}$ is the re-encrypted signal of the output $\uu(t)$ with the function $Q':\Z_q^\mmm\ra\Z_q^\mmm$, defined as
\begin{equation}\label{eq:reenc}
	\uu'(t):= \Enc(Q'(\Dec(\uu(t)))).
\end{equation}
Note that the operation of the encrypted controller \eqref{eq:controller_encrypted} coincides with the description of Figure~\ref{fig:reenc2}.

From now on, we discuss the effect of the ``injected errors'' of the cryptosystem, which has been seen in Proposition~\ref{prop:system_dec}.
Recall that, in Corollary~\ref{cor:scale}, the factor $\LLL$ has been used for scaling up the messages, in order to reduce the effect of the errors.
For this purpose, in fact, all the signals of the systems \eqref{eq:system_Z} and \eqref{eq:system_Z_q} have already been ``scaled up by $\LLL$, although it did nothing about the performance in the previous subsection.

Let us consider the performance of the controller \eqref{eq:controller_encrypted}; let the decryption $\Dec$ be taken to both sides of \eqref{eq:controller_encrypted}.
Analogously to Proposition~\ref{prop:system_dec}, the decrypted state $\tilde \zzz(t):=\Dec(\zz(t))\in\Z_q^\ell$ and the decrypted output $\tilde \uuu(t) := \Dec(\uu(t))$ will obey
\begin{align}\label{eq:controller_dec}
	\begin{split}
	\tilde \zzz(t+1) &= \FFF\cdot \tilde \zzz(t) + \GGG\cdot (Q(y(t))+\Delta_\yyy(t)) + \RRR \cdot (Q'(\tilde \uuu(t))+\Delta_{\uuu'}(t))+\Delta_\zzz(t) \mod q\\
	\tilde \uuu(t) &= \HHH\cdot \tilde \zzz(t) + \JJJ (Q(y(t))+\Delta_\yyy(t))+\Delta_\uuu(t)\mod q\\
	\tilde \zzz(0)&= \zzz_0 + \Delta_0\mod q
\end{split}
\end{align}
with some $\Delta_\yyy(t)\in\Z^\ppp$, $\Delta_{\uuu'}(t)\in\Z^\mmm$, $\Delta_\zzz(t)\in\Z^\ell$, $\Delta_\uuu(t)\in\Z^\mmm$, and $\Delta_0\in\Z^\mmm$ such that
\begin{equation*}
	\| \Delta_\yyy(t)\| \le n_0\sigma,\qquad
	\| \Delta_{\uuu'}(t)\| \le n_0\sigma,\qquad
	\|
	 \Delta_\zzz(t)\|\le (\ell + \ppp)\Delta_\Mult,\qquad \|\Delta_\uuu(t) \| \le (\ell + \ppp)\Delta_\Mult,\qquad \|\Delta_0 \| \le n_0\sigma.
\end{equation*}

To make use of the results of Theorem~\ref{thm:1} and Lemma~\ref{lem:Z}, we consider an auxiliary system defined over $\Z$, as
\begin{align}\label{eq:controller_dec_Z}
	\begin{split}
		\ol z(t+1) &= \ol F\cdot \ol z(t) + \ol G\cdot (\ol y(t))+\Delta_\yyy(t)) + \ol R \cdot \left(\LLL\cdot \left\lceil\frac{\sss^2}{\LLL}\cdot \ol u(t)\right\rfloor+\Delta_{\uuu'}(t)\right)+\Delta_\zzz(t)\\
		\ol u(t) &= \ol H\cdot \ol z(t) + \JJJ (\ol y(t)+\Delta_\yyy(t))+\Delta_\uuu(t)\\
		\ol z(0)&= \ol z_0 + \Delta_0.
	\end{split}
\end{align}
Then, analogously to Lemma~\ref{lem:Z}, it can be verified that the state $\tilde x(t)$ and the output $\tilde u(t)$ defined as \eqref{eq:relation_Z} will obey
\begin{align}\label{eq:performance_dec}
	\begin{split}
	\tilde x(t+1) &= F \tilde x(t) + G y(t) + e_x'(t)\\
	\tilde u(t) &= H \tilde x(t) + Jy(t) + e_u'(t)\\
	\tilde x(0) &= e_0'
\end{split}
\end{align}
where the errors $\{e_x'(t),e_u'(t),e_0'\}$ are determined as
\begin{align*}
	e_x'(t) &=e_x(t) + \Delta_x^\LLL(t) := e_x(t) + \frac{\rrr\sss\cdot T^{-1}\cdot (\ol G \Delta_\yyy(t)+\ol R \Delta_{\uuu'}(t) + \Delta_z(t))}{\LLL}+ \frac{\rrr\sss^2\cdot TR(\ol J \Delta_\yyy(t) + \Delta_\uuu(t))}{\LLL}\\
	e_u'(t)&=e_u(t)+\Delta_u^\LLL(t) :=  e_u(t) +
	\frac{\rrr\sss^2\cdot(\ol J \Delta_\yyy(t)+\Delta_\uuu(t))}{\LLL} \\
	e_0'&= e_0+ \Delta_0^\LLL := e_0 + \frac{\rrr\sss\cdot T^{-1}\cdot \Delta_0}{\LLL},
\end{align*}
where $\{e_x(t),e_u(t),e_0\}$ are the same quantization errors defined in \eqref{eq:errors} and \eqref{eq:error_z_to_x}. Note that $\{\Delta_x^\LLL(t),\Delta_u^\LLL(t),\Delta_0^\LLL\}$ are bounded as
\begin{align*}
	\max\{\|\Delta_x^\LLL(t)\|,\|\Delta_u^\LLL(t)\|,\|\Delta_0^\LLL\|\}&\le
	\frac{\rrr}{\LLL}\cdot \max\{\|T^{-1}\|,\|TR\|,1\}\cdot((\max\{\|G\|+\|R\|,\|J\|\}+1)n_0\sigma + (\ell + \ppp)\Delta_\Mult)\\
	&=:\beta(\rrr,\LLL)
\end{align*}

Now, in the following theorem, we prove that the performance of the controller \eqref{eq:controller_encrypted} is the same with that of \eqref{eq:performance_dec}, where the errors $\{e_x'(t),e_u'(t),e_0'\}$ consist of the errors $\{e_x(t),e_u(t),e_0\}$ due to quantization, and the rest due to error injection of the cryptosystem.
Since only the signals are scaled by $\LLL\in\N$ before encryption and the injected errors are not, it can be understood that the effect of the cryptosystem errors can be made arbitrarily small, by increasing the factor $\LLL$.

\begin{thm}\label{thm:2}
	Given any controller \eqref{eq:controller_given} and $\epsilon>0$,
	under Assumption~\ref{def},
	there exist $\rrr'>0$, $\sss'>0$, and $\LLL'\in\N$ such that for any $\rrr<\rrr'$, $\sss<\sss'$, and $\LLL> \LLL'$,
	the encrypted controller \eqref{eq:controller_encrypted} with \eqref{eq:q}
	ensures $\|g(\Dec(\uu(t)))-u(t)\|\le \epsilon$, $\forall t\ge0 $.
\end{thm}
{\it Sketch of Proof:}
The proof is analogous to the proof of Theorem~\ref{thm:1}.
Given $\delta(\epsilon)>0$ from Assumption~\ref{def},
analogously to Lemma~\ref{lem:Z} with \eqref{eq:alpha},
we can choose $\rrr'$, $\sss'$, and $\LLL'$
such that $$\alpha(M,M+\epsilon,M+\epsilon,\rrr',\sss')+ \beta(\rrr',\LLL')\le  \delta(\epsilon).$$
This ensures that the auxiliary system \eqref{eq:controller_dec_Z} satisfies $\|(\rrr\sss^2/\LLL)\cdot \ol u(t) -u(t)\|\le \epsilon$, provided that Assumption~\ref{def} holds.
Then, it follows that \eqref{eq:bound_Z} holds, so that
analogously to the proofs of
Proposition~\ref{prop:Z_q} and
Theorem~\ref{thm:1},
the condition \eqref{eq:q} guarantees that \eqref{eq:thm} holds for all $t\ge 0$.
It follows that $$g(\Dec(\uu(t)))= g(\tilde \zzz(t)) = \frac{\rrr\sss^2}{\LLL}\cdot \ol u(t)$$
holds for all $t\ge 0$,
so the proof is completed.\hfill$\blacksquare$

Finally, we review the implications of Theorem~\ref{thm:2}.
There have been two issues that hinder the unlimited operation of linear dynamic systems implemented over encrypted data; the first issue was due to the recursive multiplication by non-integer numbers, and it has been handled by the introduced method of converting the state matrix to integers.

Then, the second issue was the ``error growth'' problem of the cryptosystems.
Even though the given state matrix of the system consists of only integers, most homomorphic cryptosystems (that allow for both the addition and multiplication) do not support the recursive multiplication by the encrypted numbers for an infinite number of times, unless bootstrapping techniques of fully homomorphic encryption is utilized.
And, this is because of the error growth problem; once the injected errors exceed a certain bound, then the correct computation outcome cannot be expected.

To allow for the unlimited recursive multiplication,
the benefit of GSW-LWE cryptosystem has been first discussed.
Considering that linear systems only update the state (multiplicand) and does not update the matrix (multiplier), it has been observed that the matrix encrypted using GSW scheme can be multiplied to the LWE-based encrypted vectors for an infinite number of times, thanks to the structure of LWE-type ciphertexts where the errors are stored together with the messages.
And, despite the recursive multiplication repeated unlimited times, it has been shown that the growth of the injected errors does not tend to be infinitely large, and its effect is suppressed and controlled, under stability.
The effect of the errors can be identified with perturbations (or external disturbances), and its size can be made small by appropriate choice of parameters, so that it can keep the  performance as practically equivalent to the given un-encrypted model.

Compared to the methods from cryptography utilizing the bootstrapping techniques of fully homomorphic encryptions, the introduced method  exploits addition and multiplication over ciphertexts only, which does not require a substantial amount of computational resources.
A couple of related results are introduced; a guideline of choosing the cryptosystem parameters $(n,q,\sigma_0)$ and the control parameters $(\rrr,\sss,\LLL)$ that guarantees both the desired level of security and control performance is found in \cite{kim2020design}.
And, further illustrative explanations using MATLAB example codes can be found in \cite{KSH20compre}.

\section{Conclusion}\label{sec:conclu}

We have introduced several approaches for encrypted control, based on homomorphic encryption, multi-party computation, and secret sharing. We have compared their benefits and weaknesses, and discussed the trade-offs between them.
It has been suggested that the encrypted control approach and the corresponding cryptosystem should be chosen with an engineering point of view, so that its security model, enabled operations, and computation efficiency is suitable for the system under construction.

Among the research directions struggling to overcome the trade-offs between security, computation efficiency, and the operation ability, we have introduced a homomorphic encryption based problem which aims for implementing linear dynamic systems exploiting only addition and multiplication over integers.
Then, it has been addressed that the problems from cryptography can be tackled with the control perspectives; we have addressed that the error growth of the GSW-LWE cryptosystem can be regarded as perturbations in control systems, and the issue of incapability of recursive multiplication by non-integer state matrix has been resolved using a method based on pole-placement technique.

\section{Acknowledgement}
This study was supported in part by the Research Program funded by the SeoulTech(Seoul National University of Science and Technology), and in part by the Knut and Alice Wallenberg Foundation.



%% References


\bibliographystyle{elsarticle-num}
\bibliography{References}






% \end{thebibliography}

\end{document}